\documentclass[a4,12pt]{article}
\usepackage[english]{babel}
\usepackage{amssymb}
\usepackage{amsmath}
\usepackage{authblk}
\input{epsf}

\newcommand{\be}{\begin{equation}}
\newcommand{\ee}{\end{equation}}
\newcommand{\diag}{\rm diag}

\begin{document}

\title{Formation of the large-scale structure of the Universe}

\author{V. N. Lukash, E. V. Mikheeva, A. M. Malinovsky\thanks{Electronic address: \texttt{Alexandr.M.Malinovsky@gmail.com}}}
\affil{\it Astro Space Center of Lebedev Physical Institute, Russian
Academy of Sciences, Profsoyuznaya st. 84/32, Moscow, 117997,
Russia}

\date{}

\maketitle

\begin{abstract}
In this review, the formation, evolution, and decay of the
large-scale structure of the Universe is discussed in the context of
observational data, numerical simulations, and the Cosmological
Standard Model (CSM). Problems concerning measuring and interpreting
cosmological parameters, determining the composition of matter, and
normalizing density perturbation spectra are especially highlighted.
\end{abstract}

\tableofcontents

\section{Indroduction}

This review describes the processes of generation, evolution, and
degradation of structures -- linear, quasilinear, and nonlinear --
and relaxed halos of dark matter (DM) in the Universe, based on
theoretical developments and comparisons with observations and
numerical simulations. Special attention is paid to: (1) the
characteristics of dark energy (DE) and methods of measuring it, (2)
DE, which gravitationally affects the growth of cosmological
inhomogeneities and drastically influences the dynamics of DM
structure formation, and (3) questions of normalizing density
perturbation spectra on the basis of observations of the Universe's
large-scale structure. We separately consider the current status of
DM equilibrium halos (their internal structure, density
distribution, rotation curves, etc.) in
review~\cite{doroshkevich2012}.

Turning to the most challenging questions of physical cosmology, we
do not attempt to highlight all aspects related to the formation and
decay of large-scale structure in the Standard Model (in particular,
we do not elaborate on the baryon history of the structure). Both
reviews rely on the original research of the authors and follow
respective chapters of monograph~\cite{lukash2010}. Here, we propose
a more complete account of the aforementioned topics, invoking new
observations and comparisons to the theory, and avoiding detailed
analytical manipulations whenever possible (all necessary
theoretical results are formulated in the Appendices; the interested
reader will find proofs in book~\cite{lukash2010}).

\section{What the structure is}

The Hubble flow (recession, rushing outward, expansion) of matter
observed on large scales bears \textit{no structure}: it is
compatible with spatially homogeneous and isotropic field of
distributions of density, pressure, velocities, and other
characteristics of matter. The structure is, by definition, \textit{
inhomogeneous} and is associated with distortions of the Hubble flow
that have evolved from initial seeding inhomogeneities of the flow
metric under the action of gravitational field \textit{gradients},
which influence the motion and distribution of matter in space (see
Appendices A and B). Spatial gradients grow if scales decrease.
Relatedly, one distinguishes between linear, quasilinear, and
nonlinear structures.

By the large-scale structure of the Universe is meant the observed
inhomogeneous matter distribution which deterministically evolved
from the initial small geometric scalar perturbations (the S-mode of
cosmological perturbations) `imprinted' in the gravitational
potential of the Hubble flow.\footnote{These perturbations are also
called adiabatic or growing adiabatic density perturbations.
Theoretically, one may also conceive of primary perturbations coming
from inhomogeneities in the matter composition under invariable
gravitational potential (so-called isometric perturbations). There
are, however, no observational data (within the arrow-bars of the
measurement) which indicate that a part of the initial conditions
might be described by these perturbations.} The modern Universe
exhibits a well-developed non-linear structure in the form of halos
of galaxies, groups, and clusters at small scales ($R < 10$ Mpc),
and shows a more regular, quasilinear distribution of matter on
larger scales up to hundreds of megaparsecs, exemplified by
superclusters and cosmological voids. There are numerous transition
forms between nonlinear and quasilinear structures.

Gravitationally bound halos are distributed nonuniformly in the
Universe. They are frequently observed in sheet-like formations --
`walls' whose transverse size does not exceed 10 Mpc. A wall does
not expand transversely and appears to be a nonlinear formation. It
can be quasilinear along the other two more extended directions,
continuing to expand along them. The walls themselves are
inhomogeneous, and appear as a collection of elongated `filaments'
which may intersect, forming `nodes' -- the rich clusters of
galaxies. The mean inhomogeneity scale in the Universe measures 10
Mpc (the density contrast variance in a sphere of this radius is
close to 1, $\sigma_{10}\simeq 1$), varying locally by increasing in
the vicinity of clusters, and decreasing away from them.

The observed structure is associated with the development of
gravitational instability in nonrelativistic collisionless matter.
Initial quasiisotropic expansion of matter is accompanied by the
development of anisotropy. In regions of augmented density, the
expansion of gravitating matter slows down, comes to rest, and is
superseded by a collapse. The initial stage of collapse proceeds
mainly along one of three directions and leads to self-crossings
(caustics) and the formation of one-dimensional oppositely directed
flows. Later on, the regions of matter self-confined by its
gravitational field relax, gradually acquiring a spherical shape and
forming multistream systems trapped by gravity -- the halos of DM.
These processes, both well understood and studied, are corroborated
by numerical experiments simulating billions of collisionless
gravitating particles (the $N$-body simulations), and agree largely
with observations (see, for example, Refs~\cite{tinker2010,
huffenberger2003}).

Quasilinear structures (the density contrast $\delta_R < 1$ [see Eqn
(62) in Section 13]) are subject to a rigorous analytical treatment
since there is a small parameter enabling the machinery of
perturbation theory in this case. There are two analytical
approximations describing nonlinear formation -- that of Zel'dovich
[5] (exact in the one-dimensional case), and Press-Schechter [6]
(exact in the spherical case). As indicated by numerical
simulations, the Zel'dovich approximation describes well the
large-scale matter distribution in regions where the collapse has
just begun. In contrast, the Press-Schechter formalism pertains to
fairly small scales and describes the distribution of virialized
halos of DM.

\newpage

\section{Why galaxies form}

We observe the state of the Universe billions of years after the Big
Bang. On large scales, the expansion of matter follows Hubble's law,
which does not make a distinction between spatial points of the
medium because the relative recessional velocity $\delta\mathbf V$
of {\it any} two neighboring elements of matter is proportional to
the proper distance $\delta\mathbf r$ between them:
\be\label{huble}
\delta{\mathbf V}\equiv\frac{\partial\delta{\mathbf r}}{\partial t}
= H\delta{\mathbf r}\,.
\ee
The proportionality coefficient $H = H(t)$ does not depend on the
spatial coordinates or mutual location of these
elements,\footnote{The modern value is $H_0 \simeq 70$
km$\,\mbox{s}^{-1}\mbox{Mpc}^{-1}$.} but depends on the proper
physical time $t$. This law of matter expansion is preserved as a
relict from the early history of the Universe, being, in essence,
synonym of the notion of the Universe. It conceals all the
information on the formation of the Universe and the seeds in its
structure.

Even though $H$ is independent of spatial coordinates $x$, the
metric described by Eqn (\ref{huble}) is homogeneous and isotropic
only {\it locally},\footnote{Summing up vectors of distances and
velocities on the hypersurface $t = const$, one readily concludes
that if law (\ref{huble}) holds true for a particular observer, this
same law is then also valid for all other points with just the same
coefficient $H$. This, however, {\it does not prove} the global
homogeneity of space: we are dealing with a tautology, because an
isotropic and homogeneous hypersurface is assumed {\it a priori} in
this conjecture in the form of linear superposition of distances and
velocities. Notice that the dependence of $H$ on $\mathbf x$ can
manifest itself on scales in excess of the size of the observable
Universe. The real scale of the Friedmann world -- the region of
homogeneity where the linear superposition of velocities is valid --
exceeds the Hubble radius [see Eqn (7) below]. Finding it is a
question of the accuracy of observational data.} i.e., it still
depends on ${\mathbf x}= x^i\, (i=1,2,3)$. The Hubble flow
(\ref{huble}) can be conceived of geometrically as a
three-dimensional spatial (initially curved) hypersurface that
uniformly stretches with time, preserving local isotropy at all
spatial locations, with the factor of local stretching
$a(t)\!\cdot\mathfrak{a}_{ij}({\mathbf x})$\,, where $a=a(t)$ and
$\mathfrak{a}_{ij}=\mathfrak{a}_{ij}(\mathbf x)$ are some smooth
continuous functions of class $C^{\,2}$ .

Indeed let us assume by definition
$$
H_i^j\equiv\dot a_{ik}\, a^{kj}\,,
$$
where $a_{ij}$ and $a^{ij}$ are direct and inverse symmetric
positive definite matrices (the dot over a letter implies a partial
derivative with respect to $t$). From Eqn (\ref{huble}) we
understand that they admit the factorization
\be\label{zh}
H_i^j=H\delta_i^j\,:\;\qquad a_{ij}=
a\cdot\mathfrak{a}_{ij}\,,\;\quad a= \exp\!\left(\int\!
H\,dt\right),
\ee
and correspond to the interval squared between medium points:
\be\label{qf}
ds^{\,2}=dt^{\,2}-\delta{\mathbf r}^{\,2}= dt^2 - a^2g_{ij}\,d x^i d
x^j\,,
\ee
$$
\delta{\mathbf r}=\delta r_i\equiv
a\cdot\mathfrak{a}_{ij}\,dx^j\,,\qquad g_{ij}=
g_{ij}\!\left(\mathbf{x}\right)\equiv\mathfrak{a}_{ik}
\mathfrak{a}_{lj} \delta^{kl}
$$
[the Kronecker symbol here follows from Eqn (\ref{zh})]. The
converse statement is also true: differentiating $\delta{\mathbf r}$
from Eqn (\ref{qf}) with respect to time we recover Eqn
(\ref{huble}) with $H=\dot{a}/a$. Relatedly, Eqns (\ref{huble}) and
(\ref{qf}) are equivalent. They describe a {\it homogeneous}
distribution of matter with a laminar Hubble flow such that the
linear law of velocity superposition holds in the vicinity of any of
its points. Passing from one spatial domain to another one, we have
to redefine the rules of distance addition with regard to functions
$g_{ij}$\,.

The modern value of the scale factor is taken to be equal to
unity:\, $a\equiv (1+z)^{-1}$, where z is the redshift. In order to
determine the boundaries of the Friedmann world, we expand $g_{ij}$
in the Taylor series in some finite neighborhood of an arbitrary
point:
\be\label{rt}
g_{ij}({\mathbf x})= c_{ij} + k_0\, c_{ijk}x^k+ \frac 12 k_0^2\,
c_{ijkl} x^k x^l+\ldots,
\ee
where the coefficients $k_0$ and $c_{ij\ldots}$ depend on the
selected point, which we have placed at $\mathbf x=0$ in this case,
and $k_0^{-1}$ is the scale of variation of trace $\gamma_{ii}$ or
the convergence radius $k_0\vert\mathbf x\vert <1$ within which the
first term $c_{ij}$ of the series exceeds the remaining sum. Since
we are dealing with the form $g_{ij}dx^idx^j$, we can always reduce
$c_{ij}$ to the unity form $c_{ij}=\delta_{ij}$ by an appropriate
choice of coordinates and leave only nonvanishing coefficients
$c_{ijk\ldots}$. Expression (4) reduces then to the following:
\be\label{rt1}
g_{ij}(\mathbf x)=e^{-2q_0}\left(\delta_{ij}- 2{\mathfrak{S}}_{ij}
\right),
\ee
where the irreducible scalar $q_0=q(\mathbf x)$ and small tensor
$\mbox{\boldmath $\mathfrak S$}_0=\mathfrak{S}_{ij}(\mathbf x)$
depend on $\mathbf x$ and are close to zero at ${\mathbf x}\sim 0$
(the subscript `0' on these functions implies the absence of the
functional dependence on time).\footnote{To be fully rigorous, we
have to add to the right-hand side of Eqn (5) some scalar $B_{,ij}$
and vector $\xi_{i,j}$ terms depending on $\mathbf x$ (the comma in
subscripts denotes a partial derivative over $\mathbf x$). These
terms, however, lack physical sense and can be removed through
coordinate transformations. Vector modes, as well as decaying
branches of scalar (S) and tensor (T) perturbation modes, are
incompatible with isotropic expansion (2). In contrast, the growing
branches of S and T modes of geometric inhomogeneities are preserved
in the form (3) on large scales and do not violate Hubble's
expansion law (2). Actually, these `frozen' gravitational potentials
$q_0$ and $\mbox{\boldmath $\mathfrak S$}_0$ constitute the seeds of
cosmological structure. The decaying modes had already faded out
before the formation of the Hubble flow began and certainly had been
negligibly small by the beginning of the galaxy formation epoch.}

According to Eqn (5), the quantity $q(\mathbf x)$ is defined up to
an additive constant, and the difference in its values between
distant points can be arbitrarily large. It is usually assumed that
its mean value $q_0$ in the observed domain equals zero. In that
case, the deviations of $q(\mathbf x)$ from zero grow with the
distance from the observer, and the upper bound on the Friedmann
world is set by the condition
\be\label{q00}
|q(k_0^{-1})|\sim 1\,.
\ee
This size is {\it a fortiori} larger that the radius of external
curvature ($k_0 < H_0$), since $\vert q(\mathbf x)\vert\ll 1$ within
the scale of observable cosmology:\footnote{To derive relationship
(7), one needs to make use of the Sachs-Wolfe
formula~\cite{sachs1967}
$$
\delta_{TSW}= q_{HZ}/5\simeq 10^{-5}
$$
and estimate $q_H$ with the help of spectral integral
$$
\langle q^2_H\rangle = \int_{H_0}^\infty q_k^2\frac{dk}{k}\sim
q_{HZ}^2\,\ln\!\left(\frac{k_{\rm eq}}{H_0}\right) \simeq (2\,
q_{HZ})^2\simeq 10^{-8}\,,
$$
where $q_{HZ}=q_{k\sim H_0}$ is the large-scale Harrison-Zeldovich
spectrum~\cite{harrison1970, zeldovich1972}, and $k_{\rm
eq}=\dot{a}_{\rm eq}\simeq 0.01$ Mpc$^{-1}$ is the Hubble scale at
the instant of time when radiation and matter densities are equal.
Notice that relationship (6) is valid under the assumption of
$|\mbox{\boldmath $\mathfrak S$}_0(k_0^{-1})|\le 1$. The function
$\mbox{\boldmath $\mathfrak S$}_0$ carries information on
cosmological gravitational waves and partly on the pre-inflational
geometry of spacetime (the other part is hidden in the function
$q_0$). It is not, however, related to galaxies and, according to
observations, is much smaller than $q_0$. We do not consider it in
detail for these reasons.}
\be\label{q01}
q_H=|q(H^{-1}_0)|\sim 10^{-4}\,.
\ee

Functions (5) encode information on the \,S and T modes of
cosmological structure, which define the anisotropy of cosmic
microwave background radiation ($q_0$ and $\mbox{\boldmath
$\mathfrak S$}_0$) and the seeds of galaxies (only $q_0$). It should
be borne in mind that there are small corrections $\sim
\mbox{\boldmath $\nabla$}q_0/{\bar H}$ to the metric (3) and that in
this order the Hubble flows are weakly distorted. The gradients of
$q_0$ may evolve and disrupt the laminar flow at small scales,
leading to its breakup into self-gravitating nonlinear clumps of
matter.

We see that the early Universe is {\it deterministic} and all its
motions can be considered in the framework of the Cauchy problem. By
solving dynamical equations, we uncover the cause-consequence chain
of events fully determined by the initial cosmological conditions
(functions $q_0$ and $\mbox{\boldmath $\mathfrak S$}_0$), which in
fact gives rise to the cosmological time arrow.\footnote{The time
arrow can be violated in certain regions of spacetime where the
relativistic effects are important (for example, in black holes or
wormholes).}

In summary, there are two regimes of matter organization, which are
evolutionarily connected with each other and describe opposite
processes: Hubble flows (large scales), and structure (small
scales). From the GRT equations it follows that the metric (3)-(5)
represents the leading term in the expansion of the \textit{exact}
solution in the small parameter
\be\label{po91}
\frac{\beta k}{\bar H} < 1 \,,
\ee
where $\beta$ is the mean speed of sound in the medium, and $k$ and
 $\bar{H}\equiv aH=\dot a$ are the spatial and Hubble frequencies,
respectively. Of principal importance is the answer to the question
of whether corrections to expression (3) grow or decay with time.
The evolution leads to the \textit{breakup} of existing Hubble flows
in the first case, and to their \textit{creation} in the latter.

The answer to this question depends on the sign of function
$\bar\gamma$, where
\be\label{gamma}
\bar\gamma\equiv-\frac{\dot{\bar H}}{\bar{H}^{\,2}} =-\frac{d\ln
\bar{H}}{d\ln a}=\frac 12\left(1+\frac{3p}{\varepsilon}\right),
\ee
\[
\gamma\equiv- \frac{\dot H}{H^2} =1+\bar\gamma=\frac 12\left(1+
\frac{p}{\varepsilon}\right).
\]
Indeed, in the first order in gradients of $q_0$ we
have~\cite{lukash2010}
\be\label{qu}
q= q_0+ \int\!\delta_p\frac{da}{a}\,,
\ee
\be\label{g}
{\mathbf v}_{\rm pec}=-\nu\mbox{\boldmath $\nabla$}q_0\,,\qquad
\delta = g\Delta q_0\,,\qquad \Phi=\phi q_0\,,
\ee
where $q=q(t,{\mathbf x})$ is the curvature
potential~\cite{lukash1980a, lukash1980b}, $\phi$ is the
gravitational potential, ${\mathbf v}_{\rm pec}$ is the peculiar
velocity of matter motion relative to Hubble flow (1), and
$\delta_p$ and $\delta_\varepsilon\!\equiv\delta$ are the comoving
perturbations of total pressure ($p$) and matter energy density
($\varepsilon$) (see Appendices A and B). The growth factors
$\nu=\nu(a)$, $g=g(a)$, and $\phi=\phi(a)$ depend only on time or
the scale factor $a(t)$ respectively:
\be\label{fros}
\nu=\frac{1}{a^2}\int\frac{da}{H}\,,\qquad \phi=\gamma \bar H^2 g= 1
- \frac{H}{a} \int \frac{da}{H}\,.
\ee

A critical regime for the growth functions is the linear
cosmological expansion for which
$$
\bar\gamma=0\,,\qquad a\propto t\,.
$$
In this case, $\nu$, $g$, and $\phi$ are constant in time, and the
evolution regimes mentioned above are separated: the Hubble flows
are preserved at locations where they have existed, and no
generation of new structures takes place.

For a decelerated expansion ($\bar\gamma>0$), the functions $\nu(a)$
and $g(a)$ grow with time, a gravitational instability takes place,
the initially laminar medium flows become disturbed, and the
conditions arise for inhomogeneous structure formation on the side
of small wavelengths (the function H monotonically decreases).

For an accelerated expansion ($\bar\gamma< 0$), the functions $\nu$
and $g$ decay with time, and a new structure is not created but, in
contrast, the build-up of Hubble flow and function $q(\mathbf x)$
over an increasing range of scales $k_0< k <\bar H$ continues (the
function $\bar H$ monotonically grows).

One may conclude that gravity in equal degree spawns two dynamical
properties: repulsion (the generation of Hubble flows), and
attraction (the generation of structure). Which of them will prevail
depends on the equation of matter state [see Eqn (9)]: the inflation
(repulsion, the generation of the Hubble flow) is realized for
$(\varepsilon+3p)< 0$, and the deceleration (attraction and the
development of collapse) in the opposite case for $(\varepsilon+3p)>
0$. Being inherent to GRT, both inflation and collapse occur for
rather general initial distributions and properties of matter and
lead to the emergence of \textit{ordered} geometrical configurations
on various scales -- the Hubble flows, and nonlinear halos of
matter.

The convergence of integral (10) at the lower limit assumes the
dominance of initial adiabatic perturbations. If the isometric
pressure scalar is neglected, the relationship between $\delta_p$
and $\delta$ takes a simple form: $\delta_p=\beta^2 \delta$, and the
dynamics of scalar $q$ obey the independent harmonic oscillator
equation~\cite{lukash1980a, lukash1980b} (see Appendix B)
\be\label{qe}
\ddot q + \left(3H+ 2\,\frac{\dot\alpha}{\alpha}\right)\dot
q-\beta^2 \frac{\Delta q}{a^2}=
0\,,\qquad\alpha^2=\frac{\gamma}{4\pi G\beta^2}\,.
\ee
where $G$ is the Newtonian constant of gravitation. For small
velocities of sound $|\beta| < 1$, the leading solution
$q=q_0(\mathbf x)$ continues into the causally connected domain [cf.
expression (8)]. Spectral amplitudes of scalar curvature $q_0(k)$ in
the post-recombination epoch and those of initial perturbations of
the S-mode $q_k$ are linked linearly:
\be\label{T}
q_0=T\!\left(k\right)\cdot q_k\,,
\ee
where $T(k)$ is the transfer function of linear density
perturbations that accounts for their evolution in the pre-galactic
medium (for details, see Ref.~\cite{doroshkevich2012}).

\section{Boundaries of homogeneity}

In the course of inflation in the early Universe, the
\textit{stretching} of the space-like hypersurface with the size of
$k_0^{-1}$, already born by the preceding evolution, and the
building-up of function $q({\mathbf x})$ from the side of small
scales $k\gg k_0$, which enter the zone $k<\bar H(t)$ from the
microscopic region if $\dot{\bar H}>0$ took place. In this case, the
growth of scales $\propto\!a(t)$ takes over the Hubble radius
$H^{-1}(t)$ (Fig. 1), all folds and irregularities of the initial
hypersurface within $k_0^{-1}$ are smoothed out, and the newly
emerging small scales appear embedded into the already existing
large-scale framework of the hypersurface being isotropizied. We can
say that in the course of accelerated expansion of matter the
minimum comoving scale of the Hubble flow decreases.

\begin{figure}[bt]
\epsfxsize=112 mm \centerline{\epsfbox{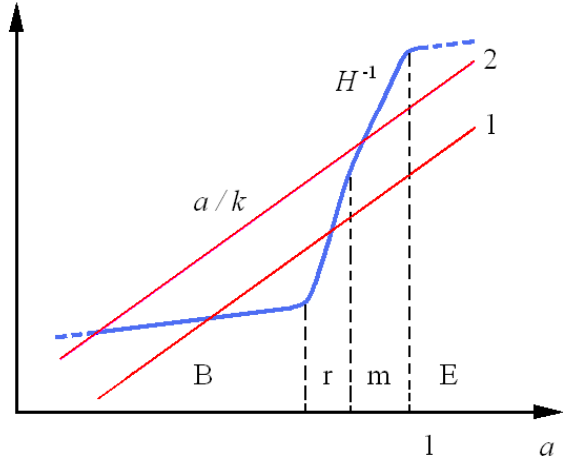}}
\caption{The Hubble radius $H^{-1}$ (the line with bends) and sizes
of perturbations $\propto a$ (inclined straight lines: 1 -- a
galaxy, 2 -- a supercluster) as functions of $a$, from the Big Bang
(B) to periods of radiation (r), DM(m), and DE (E).}
\label{goos}
\end{figure}

The maximum size of Friedmann hypersurface $k_0^{-1}$ is determined
by the conditions at the beginning of inflation and, as follows from
observations, exceeds the modern horizon ($k_0 < 2\cdot 10^{-4}\;
\mbox{Mpc}^{-1}$,\, see Eqns (6) and (7)). The minimum size $k_{\rm
m}^{-1}$ is linked to the end of the inflationary period of the Big
Bang; it is certainly less than the size of the observable structure
and compares to the wavelength of the background radiation quantum.
In the process of post-inflationary decelerated matter expansion,
the opposite process took place: the Hubble flow broke up from the
side of small scales and its minimum size grew with time.

In order to determine the minimum scale of current Hubble flow (1),
we write down the first terms of metric expansion in the Lagrangian
coordinates $(t_{\rm c},{\mathbf x})$\, (the label `c' of comoving
time is dropped where possible):
\be\label{ru1}
ds^{\,2}= \left(1- 2\delta_p\right)dt^{\,2} -\delta
\mathbf{r}^{\,2}\,,
\ee
\be\label{ququ}
\delta\mathbf{r}=\delta r_i = a e^{-q}\!\left(\delta_{ij}-
{\mathfrak{b}}\,q_{0,\,ij}\right)d x^j\,,
\ee
and rewrite Hubble's law in a more precise form [cf. Eqns (92) and
(93) in Appendix A]
\be
\delta V_i\equiv\frac{\partial\,\delta r_i}{\left(1-
\delta_p\right)\partial t}=H_{ij}\,\delta r^j\,,
\ee
where
\be
H_{ij}= H\left(\delta_{ij}-\bar{h}\,q_{0,\,ij}\right),\qquad
\bar{h}\equiv\frac{\nu}{\bar H}\,,\qquad {\mathfrak{b}}=\int
\bar{h}\frac{da}{a}\,.
\ee
Neglecting the effective speed of sound, we have at the lower
boundary $k_1$ of the Hubble flow:
\be
\delta\sim 1\,,\qquad {\rm v}_{\rm pec}\sim\frac{\bar
H}{k_1}\,,\qquad q_0 \sim\frac{\bar H^{\,2}}{k_1^{\,2}} \ll 1\,.
\ee

Recalling that the observed mean inhomogeneity scale $k_1\sim 0.1\;
\mbox{Mpc}^{-1}$ (Fig. 2), we obtain for the spectral amplitude of
the curvature potential at the lower boundary of the current flow:
\be
q_0\!\left(k_1\right)\sim 5\cdot 10^{-6}
\ee
in agreement with the required value of the transfer function
$T(k_1)\sim 0.1$\,.

\begin{figure}[bt]
\epsfxsize=112 mm \centerline{\epsfbox{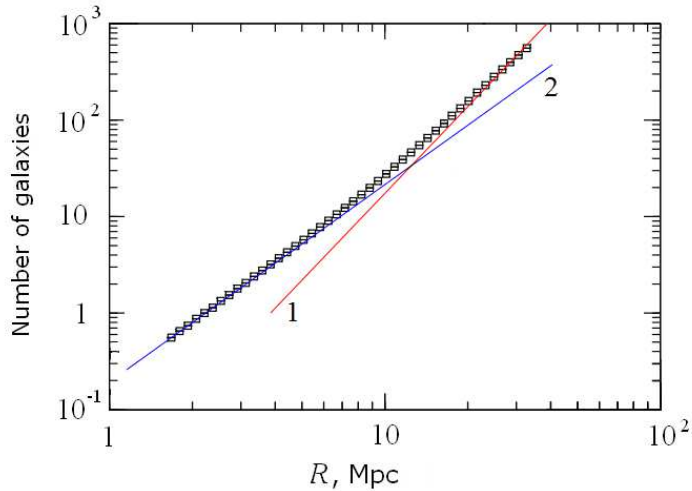}}
\caption{The spatially mean number of galaxies $N(R)$ in a sphere of
radius $R$ (according to Ref.~\cite{sylos2009}). The accuracy of
power-law asymptotics is about $10\%$: 1 -- $N \propto R^3$, the
uniform distribution of matter, and 2 -- $N \propto R^2$, the
nonlinear structure of the Universe.} \label{pilip}
\end{figure}

Of interest is the question of determining experimentally the size
$k_0^{-1}$ of the Friedmann `background' of our Universe. Doing so
is possible, in principle. In fact, the scale $k_0^{-1}$ need not be
very large, because it is connected with the \textit{last} period of
accelerated expansion in the chain of inflationary stages of the Big
Bang,\footnote{Inflationary stages could have alternated with those
dominated by matter which formed through the decay of intermediate
short-lived inflatons.} and this period could be relatively short.
In that case, $k_0^{-1}$ can, in principle, exceed the radius of
external curvature $H_0^{-1}$ only slightly. In this situation, the
total curvature potential $q_0({\mathbf x})$ generated toward the
end of inflationary explosion and determining the geometry of the
observable world would be composed of \textit{two} components: a
\textit{non-Gaussian}, strongly-correlated smooth part of size
$k_0^{-1}$, linked to global pre-inflationary geometry, and
small-scale (with respect to $k_0^{-1}$) ripples of a Gaussian field
of inhomogeneities born in a quantum-gravitational way. The
non-Gaussian component of density perturbations grows with an
increase in the distance from the observer and manifests itself in
the large-scale anisotropy of cosmic microwave back- ground. The
Gaussian perturbations do not decay as the scale is reduced, and are
responsible for the formation of galaxies.\footnote{The resultant
amplitude of quadrupole anisotropy of relic radiation can be both
lower and higher than the value expected for this extrapolation of
the short-wave spectrum. This effect could explain the low
quadrupole anisotropy of the cosmic microwave background radiation
(if, of course, this anomaly is confirmed by future observations and
rigorous data analysis).}

The post-inflationary matter-dominated stage of decelerated
expansion, accompanied by a reduction in the speed of sound, leads
to the conditions for collapse at small scales, where a `window' of
gravitational instability opens. In contrast to the inflation, where
initial conditions are forgotten, certain seed perturbations of
curvature are required for the onset of collapse. They define
domains of matter inflow and outflow.

Thus, the scale interval $ k_1 < k \ll k_{\rm m} $ in the modern
epoch belongs to nonlinear cosmological structures, and the
undisrupted quasi-Hubble flow of matter, albeit already distorted,
still persists at scales $k<k_1$ (in the mean in the Universe).

\section{The quasi-Friedmann equations}

A weakly inhomogeneous Universe is described by the generalized
Friedmann equation~\cite{lukash2010} (see Appendix A)
\be\label{qFr}
H_{\rm v}^{\,2}= \frac{8\pi G}{3}\,\varepsilon -
\frac{\varkappa}{b^2}\,,
\ee
in which the geometrical scalar variables
$$
H_{\rm v}=H_{\rm v}(t,{\mathbf x})\equiv\frac{\dot b}{b}=\frac 13
u^\mu_{\;;\mu}
$$
(the semicolon indicates the covariant derivative) and
$b=b(t,{\mathbf x})$ correspond to the local Hubble and scale
factors of medium volume expansion, $\varepsilon$ and $u^\mu$ are
the total energy density and matter four-velocity, the small
function
\be\label{qFr2}
\varkappa=\varkappa\!\left(t,{\mathbf x}\right)\equiv\frac
23\,\Delta q
\ee
is the internal space curvature, and the dot above the symbols
implies partial derivative with respect to the comoving time $t_{\rm
c}$. The time evolution of matter density obeys the conservation law
\be\label{qFr1}
\dot\epsilon + 3H_{\rm v}\left(\epsilon + p_{\rm v}\right)= 0\,,
\ee
where $p_{\rm v}$ is the volume pressure [see Eqns (77) and (85)].

Equations (21) and (23) describing the evolution of the
quasi-Friedmann Universe in geometrically invariant variables are
valid for small spatial curvature $|\varkappa|\ll 1$ and include the
zero the first orders of magnitude in deviations of weakly
inhomogeneous geometry from Friedmann's one. These equations have a
scalar form, although the geometry described by them is anisotropic
[see Eqn (89)].

In order to solve Eqn (21) in the volume factor $b$, the defining
scalar of curvature $q$ must be known. Assuming that the small
function $q$ is known to us, we seek the solution in the form
\be\label{sol}
H_{\rm v}= H_{\rm c} +\frac{\Delta\Phi-\Delta q}{3a\bar H}\,,
\ee
$$
\varepsilon = \varepsilon_{\rm c} + \frac{\Delta\Phi}{4\pi G
a^2}\,,\qquad p_{\rm v} = p_{\rm c} + \frac{\Delta S}{12\pi Ga^2}\,,
$$
where $X_{\rm c}\equiv X(t_{\rm c})$ are the background functions of
comoving time, and $\Phi$ and $S$ are arbitrary small functions of
all coordinates. Substitution into Eqns (21) and (23) results in the
\textit{correct} coupling between scalars $\Phi$ and $q$ [cf. Eqn
(86)]:
\be\label{gam41}
\Phi=\frac{H}{a}\int a\left(\gamma q - S\right)d t\,,
\ee
with the arbitrary function $S$ being not linked to the Einstein
equations.

It should be emphasized that quasi-Friedmann Eqns (21) and (23)
imply no constraints on the medium physical properties. Similarly to
the original Friedmann equations, they link the spacetime curvature
with total energy density, its time derivative, and the pressure of
matter. As concerns the physical state of the matter, it needs to be
considered only in the derivation of the equation of motion for the
curvature scalar $q$ (see Refs~\cite{lukash2010, lukash1980b} and
Appendix B).

\section{Dynamical properties of the structure}

A physical reason for the emergence of nonlinear structure -- the
formation of galaxies from small primary curvature perturbations --
is the gravitational instability of dark matter, most vigorously
developing in the post-recombinant period of DM
dominance.\footnote{It should be recalled that in the CSM the
initial perturbations in composition are absent, the background
curvature equals zero, and the parameters of the energy density of
components take the following values: DE ($\Omega_{\rm E}\approx
0.7$), nonbaryonic DM($\Omega_{\rm M}\approx 0.25$), baryons
($\Omega_{\rm b}\approx 0.05$), and radiation ($\Omega_{\rm r}\simeq
10^{-4}$). Notice that the $10\%$ accuracy level of today's
observations does not yet enable distinguishing the cosmological
constant and evolving DE. This witnesses in favor of its slow
evolution and allows considering general models of DE in the form of
expansions in terms of the small parameter $\vert w+1\vert\ll 1$,
where the cosmological constant is the leading term in the series
[see Eqn (44), $w\equiv p_{\rm E}/\rho_{\rm E}$].} Since the initial
pressure in nonrelativistic matter is low (relative particle
velocities are close to zero at any point in space), the cold medium
freely moves in own gravitational field of quasi-Hubble flow and the
initially small peculiar velocities and the contrast of matter
density grow with time on all scales. Consider at a greater length
the dynamics of the quasi-Friedmann model at the linear stage of
developing scalar inhomogeneities.

Let us turn to a simple model of the late Universe, which accounts
only for nonrelativistic matter `m' (with density inversely
proportional to the local volume, $\rho_{\rm m}\propto b^{-3}$) and
dark energy `E' of constant density ($\rho_{\rm E}=const$). In this
case, the mean speed of sound in the post-recombinant epoch is equal
to zero [see Eqn (95), \,$\beta=\delta_p=0$] and, consequently, the
scalar $q$ and spatial curvature $\varkappa$ do not depend on time,
and equation (21) simplifies. Multiplying it by \,$(b/H_{\rm E})^2$
and recasting in terms of dimensionless variables, we arrive at
\be\label{luru3}
\left(\frac{\dot b}{H_{\rm E}}\right)^2=f^{2}(b) -
\hat\varkappa(\mathbf x)\,,
\ee
where
\be\label{luru4111}
f^2(b)\equiv\frac{8\pi G\,b^2}{3H_{\rm E}^2}\left(\rho_{\rm
m}+\rho_{\rm E}\right)=\left(\frac{{\rm c_m}}{b}+
b^2\right)\,\gtrsim\, 1\,,
\ee
\be\label{ps1}
q=q_0(\mathbf{x})\equiv\frac 32\,H_{\rm E}^2 \hat
q\,,\qquad\hat\varkappa=\hat\varkappa(\mathbf{x})\equiv
\frac{\varkappa}{H_{\rm E}^2}=\Delta\,\hat{q}\,,
\ee
$H_{\rm E} = H_0\sqrt{\Omega_{\rm E}}\simeq \left( 5\;\mbox{Gpc}
\right)^{-1}$ is the Hubble constant of dark energy,\, ${\rm c_m}
\equiv \Omega_{\rm m}/\Omega_{\rm E}\simeq 0.4$ is the constant
coefficient, one sixth of which is linked to baryons and the
remaining 5/6 to DM (in this approximation, both components move
together). Obviously, the function $f(b)$ attains a minimum $f_{\rm
min}\simeq 1$ at $b_{\rm min}^{-1} \simeq 1.7$\,.

An arbitrary small function of spatial coordinates
\,$\hat\varkappa$\, describes the local normalized space curvature.
We are interested in domains with the positive right-hand side of
equation (26):
\be\label{luru4112}
\hat\varkappa\left({\mathbf x}\right)<1\,.
\ee
In these domains, the matter density decreases monotonically with
time [see Eqn (42) for details]. They comprise both superclusters
($\varkappa>0$) and cosmological voids ($\varkappa<0$).

The volume and background scale factors coincide at points
$\varkappa=0$ (the expansion anisotropy can be large in this case):
\be\label{luru5}
b=a(t)\equiv\frac{1}{1+z}\,,\qquad
H\equiv H_{\rm E}\,\frac{f(a)}{a}\,.
\ee
where $f=f(a)$ is the growth factor of the Hubble velocity component
\be\label{nhs}
\mathbf{V}_{\!H}=f H_{\rm E} \mathbf x\,.
\ee

In a general case, in the linear order in $\hat\varkappa$ we obtain
\be\label{luru6}
b=a\left(1-\frac 13\,\hat g\,\hat\varkappa\right),\qquad \delta_{\rm
m}=\hat g\,\hat\varkappa\,,
\ee
\be\label{luru7}
H_{\rm v}=H\left(1-\frac 13 \hat h\,\hat\varkappa\right),\qquad \hat
h\equiv\frac{\bar\nu}{f}=\frac{\dot{\hat g}}{H}\,,
\ee
where $\delta_{\rm m}=\delta$ is the density perturbation, $\hat
g=\hat g(a)$ and $\bar\nu=\bar\nu(a)$ are, respectively, the growth
factors of density perturbations and peculiar velocity (Fig. 3):
\be\label{luru8}
\hat g\equiv\frac{\gamma\bar H^2}{c_m}a=\frac{1}{\rm c_m}\left(a -
H\int_0 \frac{da}{H}\right),\qquad \bar\nu\equiv\frac 32 H_{\rm E}
\nu=\frac{3H_{\rm E}}{2 a^2} \int_0\frac{da}{H}\,.
\ee

\begin{center}
\begin{figure}[tb]
\epsfxsize=112 mm \epsfbox{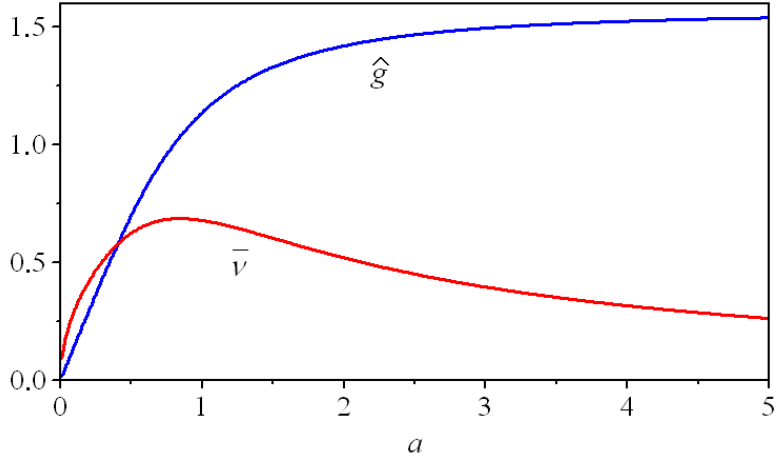}
\caption{Growth factors of density $\hat g(a)$ and peculiar velocity
$\bar\nu(a)$~\cite{lukash2008}.} \label{lururis1}
\end{figure}
\end{center}
For
\begin{align*}
a \ll 1\,&: &\hat g &\simeq \frac{3a}{2}\;, & \bar\nu &\simeq \sqrt a\,,\\
a \gg 1\,&: &\hat g &= \hat{g}_{\rm max} \simeq 1.56\;, & \bar\nu &\simeq \frac{3}{2a}\,,\\
a = 1\,&: &\hat g &\simeq 1.13\;, & \bar\nu &\simeq 0.67\,.
\end{align*}

Equations (26)-(34) describe quasi-Hubble anisotropic flows with the
effective Hubble function $H_{\rm v}$ which depends on the
observer's location. In the modern epoch, the function \,$\bar\nu$\,
shows a wide maximum, which is an indication of the period of most
intense structure generation. The position of its maximum
corresponds to $z\simeq 0.2$\,, the 90\,\% level of its maximum
value of $\bar\nu_{\rm max}\simeq 0.68$ is reached at $a\simeq 0.5$
and 1.4\,,\, and the 50\,\% level is reached at $z\simeq 0.1$ and
4\,.

The current epoch is therefore that of \textit{maximum} peculiar
velocities, which will persist for the cosmological
time~\cite{lukash2008}. The function $\bar\nu$ will decrease twofold
when the age of the Universe reaches 35 billion years. Only then
will it be possible to speculate about the beginning of the epoch of
the fading out of peculiar velocities in all space domains where
$\hat\varkappa<1$\,.

\begin{figure}[tb]
\epsfxsize=112 mm \centerline{\epsfbox{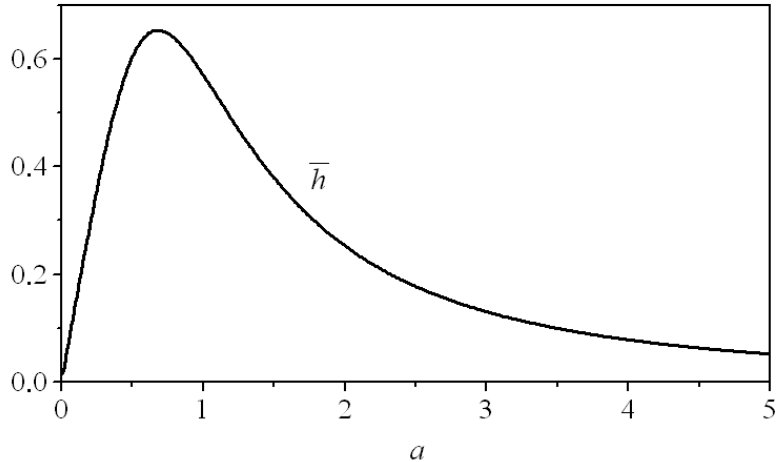}}
\caption{Function $\hat{h}(a)$ setting the distortion of the Hubble
flow~\cite{lukash2008}.} \label{lururis2}
\end{figure}

The function $\hat{h}\!\left(a\right)$ determining the deviation of
the local Hubble factor from its background value is plotted in Fig.
4. The maximum $\hat{h}_{\,\rm max}\simeq 0.65$\, is reached at
 $z\simeq 0.4$\,, while the interval of the values of $\hat{h} > \hat{h}_{\rm
max}/2$\, is within the limits \,$a \in \left( 0.1;\,1.8\right)$,
which corresponds to the range of the Universe's age, from 0.6 to 22
billion years. Figure 4 explicitly demonstrates that our Universe is
at the stage where Hubble's expansion law is maximally distorted
[$\hat{h}(a=1)\simeq 0.87\cdot\hat{h}_{\rm max}\simeq 0.57$], and
that the recovery of Hubble flows in quasilinear domains of space
will take about ten billion years.

We can conclude that the formation of large-scale structure in the
Universe lasts for the period from 1 to 22 billion years after the
Big Bang. The stage of suppression of the Hubble flow
inhomogeneities, caused by the gravitational influence of DE, has
not begun yet, although dark energy has been dominating in the
density of matter for 3.5 billion years already. The delay hinges on
the fact that the dynamic influence of DE on the structure
generation in the Universe has just begun. Indeed, the
characteristic time of this influence exceeds the current age of the
Universe and amounts to $H^{-1}_{\rm E}\simeq 17$ billion years.

\newpage

\section{Anisotropic cold flows}

The field of peculiar velocities can conveniently be described by
passing to the Eulerian coordinates in which the gravitational field
is locally isotropic in a linear order in $\varkappa$ at any space
point (see Appendix B). The relationship between the Eulerian
($\mathbf y$) and Lagrangian ($\mathbf x$) coordinates we are
interested in has the form
\be\label{luru90}
\mathbf{y}=\mathbf{x} + \hat g\,\hat{\mathbf S}\,,
\ee
where $\hat{\mathbf S}=\hat{\mathbf S}({\mathbf x})=-\mbox{\boldmath
$\nabla$}\,\hat q$ is the vector of displacement for an element of
the medium relative to its unperturbed position, $\hat\varkappa=-\,
{\rm div}\,\hat{\mathbf S}=\Delta\,\hat q$\,. The Lagrangian
coordinate $\mathbf x$ does not vary here with time along the
trajectory of the medium element and coincides with the Eulerian
coordinate $\mathbf y$, as $t\rightarrow 0$. The growth factor for
the displacement, $\hat g=\hat{g}(a)$, is simultaneously that for
the density perturbation (32). The fact that the matter displacement
relative to the laminar Hubble flow is factorized as a product of
two functions (one dependent on time, and the other dependent on
space coordinates) indicates that the rate of perturbation growth is
the same for all wavelengths.

The net displacement of medium elements with respect to the
unperturbed Hubble positions grows monotonically with time and
amounts today to 14 Mpc. The mean net displacement will approach its
limit value of about 22 Mpc (see Fig. 3) in the future, provided
that the DE density does not change.

Using equation (35), we obtain the following representations for the
interval on the Eulerian and Lagrangian grids:
\be\label{luru10}
ds^2=\left(1+2\Phi\right)d\tau^2-{\mathfrak a}^2d{\mathbf y}^2=dt^2
- {\mathfrak a}^2\!\left(\delta_{ij}-2 \hat
g\,\hat{q}_{\,,ij}\right)dx^idx^j\,,
\ee
where $\tau=t-a\,\bar\nu\, H_{\rm E}\, \hat q$, and $\mathfrak
a\equiv a(t)\cdot\!\left(1-q\right) =
a(\tau)\cdot\!\left(1-\Phi\right)$ is the local scale
factor~\cite{lukash2010}. The function $b(t,\mathbf x)$ [see Eqns
(21) and (32)] is proportional to the trace of the spatial part of
the Lagrangian metric tensor, and the gravitational potential of
density perturbations is equal to $\Phi=0.6\cdot \bar\phi\,q$, where
\be
\bar\phi=\frac{{5\,\rm c_m}}{3\,a\,}\,\hat g=\frac
53\!\left(1-\frac{H}{a}\int\frac{da}{H}\right).
\ee
For $a\ll 1$, we get $\bar\phi=1\,$. The temporal factor $\bar\phi=
\bar\phi(a)$ of the gravitational potential decay under the action
of DE is plotted in Fig. 5. The magnitude of $\bar\phi$ can serve as
a \textit{measure} of DE dynamic influence on the structure
generation. By the current epoch, the potential $\bar\phi(a=1)\simeq
0.77$ has already dropped by 23\% off its constant value in the
matter-dominated phase.

\begin{figure}[tb]
\epsfxsize=112 mm \centerline{\epsfbox{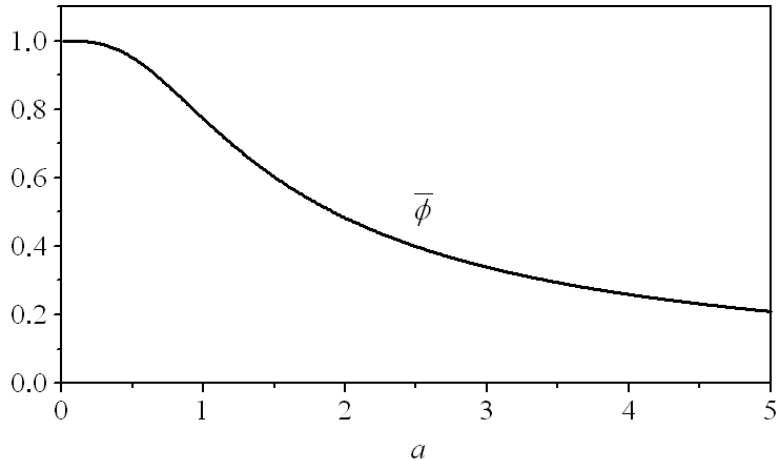}}
\caption{Decay of the gravitational potential $\bar\phi(a)$ of the
large-scale structure of the Universe under the action of dark
energy (see Ref.~\cite{lukash2010}).} \label{raspad}
\end{figure}

Interval (36) determines the physical Eulerian coordinate of medium
element $\mathbf{r}=\mathfrak a \mathbf{y}\simeq a\mathbf y$ (cf.
Eqn (105)). Differentiating $\mathbf{r}$ with respect to the proper
time, we arrive at the following formula for the peculiar matter
velocity:
\be\label{luru11}
\mathbf{v}_{\rm pec}\equiv \dot{\mathbf{r}}-H \mathbf{r}=
\hat{g}^{\,\prime}\,\hat{\mathbf S}=\bar\nu H_{\rm E}\, \hat{\mathbf
S}\,.
\ee
Expression (38) coincides with the definition of 3-velocity as the
spatial component of the 4-velocity of matter in the Eulerian
reference frame:
$$
\mathbf{v}_{\rm pec}=-\frac{\partial t}{a\partial\mathbf y}=
-\bar\nu H_{\rm E}\mbox{\boldmath $\nabla$}{\hat q}=
-\frac{{\boldsymbol\nabla}\mathfrak q}{\bar H}\,.
$$
Thus, the quantity \,$\bar\nu$\, featuring in Eqn (33) is indeed the
growth factor for the peculiar velocity.

According to Eqn (38), the total velocity of matter is defined as
$$
\mathbf{V}=H\mathbf{r}+\mathbf{v}_{\rm pec}\,.
$$
The first component $H\mathbf r$ describes the Hubble velocity of
the \textit{nonuniform} Universe. It can also be split into two
parts. One of them, $\mathbf{V}_{\!H}\!=\bar H\,\mathbf x$ [see Eqn
(31)], is connected with the homogeneous component of density, and
the other, \,$\hat g\, \bar H\,\hat{\mathbf S}$, is connected with
the perturbed one. The growth factor for the second component is
proportional to the growth factor $\hat g$ for density
perturbations, whereas the peculiar velocity is
$\propto\,\hat{g}^{\,\prime}$.

We turn now to local flows of matter. In regions of the
inhomogeneous Universe constrained by the condition (29), the flows
of matter are described by the tensor field $H_{ij}=H_{ij}(t,\mathbf
x)$ generalizing the function $H(t)$ of the Friedmann model. Indeed,
as follows from relationship (35), the coordinate distance between
nearby points of the medium at time moment $t$ is given by
\be\label{luru12}
\delta y_i = \left(\delta_{ij}-\hat g\,\hat{q}_{\,,ij}\right) \delta
x^j\,.
\ee
Differentiating the physical distance $\delta\mathbf{r}=\mathfrak
a\,\delta\mathbf y$ over time, we obtain the field of pairwise
velocities:
\be\label{luru13}
\delta V_i\equiv\frac{\partial}{\partial t}\!\left(\delta
r_i\right)=H_{ij}\,\delta r^j\,,
\ee
$$
H_{ij}=H\delta_{ij}-\dot{\hat g}\,\hat q_{\,,ij}=
H\left(\delta_{ij}-\hat{h}\,\hat{q}_{\,,ij}\right).
$$

The trace of the tensor field represents the volume Hubble function
$H_{\rm v}=H_{ii}/\,3$\, [see Eqn (33)]; however, the tensor
\,$H_{ij}$\, itself is strongly anisotropic. The anisotropy of local
expansion (variations of projections of $H_{ij}$ onto the radial
directions emanating from the given point $\mathbf x$) has the same
order of magnitude as the deviations of $H_{\rm v}$ from the mean
value of the Hubble parameter $H$. At the boundary of quasilinear
regions (29), these variations reach 100\% (up to a point of stopped
expansion in some directions). For example, in the vicinity of the
Local Group, at a distance in excess of 2 Mpc from its barycenter,
the principal values of the Hubble tensor are
$H_{ij}={\diag}(\,48\,,\, 62\,,\, 81\,)\,\mbox{km}\,\mbox{c}^{-1}
\mbox{Mpc}^{-1}$ (Fig. 6).

The field $H_{ij}$ describes regular cold flows of matter. It is
noteworthy that Eqn (40) is valid under the assumption that the
distance between galaxies is small, below the correlation radius of
the two-point correlation function for the displacement vector. For
different projections of this vector relative to the direction of
$\delta\mathbf{y}$, the correlation radius varies from 15 to 50 Mpc.
The deviations from the velocity field (40) grow with distance. The
spectrum of cosmological velocity perturbations is shaped namely in
this manner: it decays toward shorter wavelengths for $k
> 10^{-2}\,\mbox{Mpc}^{-1}$ (see, for detail, Ref.~\cite{doroshkevich2012}). For this
reason, random deviations from mean velocities (40) in this range
grow with an increase in scale.

\begin{figure}[tb]
\epsfxsize=112 mm \centerline{\epsfbox{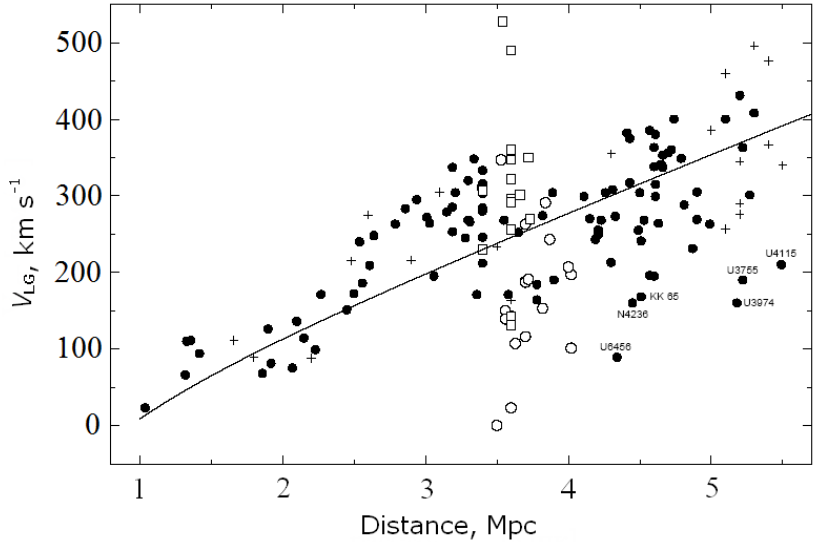}}
\caption{Radial velocities $V(r)$ of galaxies in the vicinity of the
Local Group according to data of Ref.~\cite{karachentsev2003}.
Different symbols correspond to different galaxy types, and
distances to galaxies are given relative to the barycenter of the
Local Group.} \label{karachenzev}
\end{figure}

It should be kept in mind that deviations from dependence (40)
amount to $\sim 40\,\mbox{km}\,\mbox{c}^{-1}$, i.e., about 25\% of
the mean velocity (see Fig. 6) at a distance of $\sim 2.5$ Mpc from
the barycenter of the Local Group. At the same time, the total
peculiar velocity of the Local Group relative to the cosmic
microwave background radiation comprises
$600\,\mbox{km}\,\mbox{c}^{-1}$. The scales of inhomogeneities
responsible for so high velocities are in the interval from 15 to 70
Mpc.

We see that the standard theory of the Universe's structure
formation encounters no difficulties in explaining the observed
motions of matter in quasilinear regions of the Universe
($\hat\varkappa<1$). The local flows are regular, smooth, and
strongly correlated. The smallness of random galaxy velocity
deviations from mean cold flow rates is explained by the shape of
the initial spectrum of spatial density perturbations. These flows
bear the quasi-Hubble character at small distances, preserving its
main features: the flows are cold and radial, and the speed of
galaxy recession is proportional to distance. However, the Hubble
`constant' depends on the observer position and the direction in
space. We have already mentioned the neighborhood of the Local Group
as an example. The principal values of the Hubble tensor $H_{ij}$ at
a distance of several megaparsecs from its barycenter are related as
3~:~4~:~ 5.

As the radius grows, the deviations of galaxy velocities from the
mean correlated flow rates increase too, beginning to saturate from
distances of about 10 Mpc, which corresponds to the minimum
correlation radius at which deviations of velocities reach the order
of magnitude of the Hubble velocity proper. The deviations do not
grow further, whereas the Hubble velocities continue to grow. At
large distances, the flow of matter approaches the ideal Hubble law
(1).

\section{How the collapsing flows form}

\begin{figure}[tb]
\epsfxsize=112 mm \centerline{\epsfbox{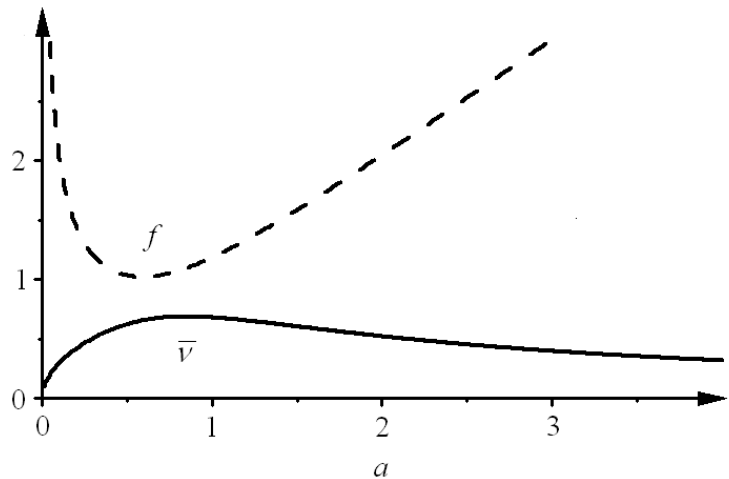}}
\caption{Growth factors of Hubble $f(a)$ (dashed line) and peculiar
$\bar{\nu}(a)$ (solid line) velocities of matter~\cite{lukash2010};
$a=1$ at $z=0$.} \label{alex4}
\end{figure}

As shown earlier, dark energy, though not a component of structure,
cardinally influences the structure generation rate and the history
of galaxy formation. Figure 7 illustrates the behavior of growth
factors of Hubble and peculiar components of matter velocity at the
quasilinear stage of evolution. Because of the peculiar velocity
growth, the expansion in certain regions of the Universe turns into
a collapse which starts in certain directions and proceeds then
toward the formation of gravitationally confined halo systems with
various masses. The spectrum of initial perturbations is shaped so
that increasingly larger masses may collapse at later time moments
(see also Ref.~\cite{doroshkevich2012}).

If DE were absent, the Hubble velocity would continue to decrease
with time, and the peculiar velocity would continue to grow. As a
result, in any spatial domain where at least one of three principal
values of tensor $\hat{q}_{\,,ij} = \diag (\,\lambda_1,\,
\lambda_2,\, \lambda_3)$ is positive [for example, $\lambda_1>0$,
where $\lambda_1\ge \lambda_2\ge\lambda_3$; see Eqn (35)], sooner or
later there will come an instant when the negative peculiar velocity
in this direction becomes equal to Hubble's one and the expansion is
halted [the principal value  $H_{11}$ turns to zero; see Eqn (40)].
In this case, the gravitational confinement of matter and its
subsequent collapse ensue. However, because of the dynamical
influence of DE not all domains succeed in evolving into the
collapse stage: the Hubble velocity reaches a minimum at $z\simeq
0.7$, and then increases sharply as $\propto a$ [see Eqn (31)],
whereas the peculiar velocity saturates and then slowly decreases
(see Fig. 7). With account for DE influence, the process of galaxy
formation resembles a fire burned out and lacking new firewood.

Thus, the nonlinear structure has a chance to form in those spatial
zones where condition (29) is violated, namely
\be
\hat\varkappa= \lambda_1+\lambda_2+\lambda_3 > 1\,.
\ee
Notwithstanding the threshold character of this inequality, it does
not imply that regions occupied with nonlinear structures are
topologically compact. Indeed,  $\lambda_3$ (or $\lambda_{2,3}$) can
appear to be negative, and then the expansion of matter will
continue in that direction (or directions).

How can the transition from the quasilinear stage to the collapse
phase be described? The answer is surprisingly simple: solution (35)
can analytically be continued up to the first self-crossing ($\delta
y_1=0$) and even further by `matching' the arising multistream flows
across caustics (the Zel'dovich approximation [5]).

This analytical approximation leads to correct qualitative
conclusions, while quantitative deviations from the real evolution
(until the first self-crossing) do not exceed 20-30\% according to
different criteria. The success of the approach, which is so simple,
hinges on the potential $q$ being small under cosmological
conditions (the formation of black holes is exponentially
suppressed). Moreover, the solution (35) proves to be exact in the
nonlinear \textit{one-dimensional} case, and namely this variant of
initial collapse development is most typical in the Universe.

Relatedly, when computing the quantitative characteristics of a
newly forming nonlinear structure -- distribution functions for
cosmological voids, superclusters, filaments, etc., mean distances
between the walls, nodes, etc., correlation functions, evolution
based on redshifts, and so on -- one can rely on linear
perturbations and their Gaussian statistics because solution (35) is
naturally set by the field of small initial density inhomogeneities.

We give below two examples. First of all, we can exactly specify
those domains where caustics fail to form, i.e., the expansion
continues forever:
\be\label{luru40}
\lambda_1 <\,\hat{g}_{\rm max}^{\;-1}\simeq 0.6\,.
\ee
As long as density perturbations remain small, the two conditions
(29) and (42) are equivalent. Condition (29), however, by no means
ensures the absence of collapse: if $\lambda_1 > 0.6$, this
direction, even if only in the future, will turn to the collapse. In
a similar way, we can refine the density field. From relationship
(35) we obtain the comoving matter density
$$
\rho_{\rm m}(\eta,\mathbf x)=\frac{{\rm c_m}\, \rho_{\rm E}}{a^3
\!\det(\delta y_i/\delta x_j)}=\frac{{\rm c_m}\, \rho_{\rm
E}}{a^3\!\left(1-\hat g\,\lambda_1\right)\left(1-\hat
g\,\lambda_2\right)\left(1- \hat g\,\lambda_3\right)}\,.
$$
Formula (32) follows obviously from the last one for $\hat g\ll 1$.

We will return to nonlinear structures in the subsequent sections,
and now proceed with the measurements of DE.

\section{How to measure dark energy}

DE can be measured only by means of observational cosmology. Its
detection in the laboratory (similarly, for example, to attempts at
laboratory detection of dark matter) seems to be implausible because
this ever-penetrating sub- stance (field) only weakly interacts with
all objects, be they interiors of stars, compact objects, the early
Universe, or others, during the entire world history known to us.
Only indirect measurement methods are possible. Luckily, they have
already proven their efficiency and led to impressive results -- the
discovery of DE proper.

What does it mean to measure DE? The answer to this question is
hinted at by cosmology. The known constraints on the main DE
parameter, viz.
\be\label{ote}
\vert1+w\vert < \,0.1
\ee
(see, for example, Ref.~\cite{komatsu2011}) point to the constancy
of function $w\equiv p_{\rm E}/\varepsilon_{\rm E}\approx const$
both in time and in space. Condition (43) simplifies the approach to
DE detection to some extent. The DE clustering effects are of small
significance and escape detection at the current level of
technological development.

Today, we can only discuss actual experiments dealing with the
dependence of function $w$ on time: $w(a)$. Taking into account the
slowness of evolution, this function can be replaced by the set of
constant coefficients $c_n$ of its Taylor expansion around $-1$\,:
\be\label{fwf}
w\!\left(a\right)=-1+c_0+c_1\Delta a+\ldots + c_n\frac{\left(\Delta
a\right)^n}{n!}+\ldots,
\ee
where $\Delta a=1-a\equiv z(1+z)^{-1}$. At the current level of
knowledge, the physical information on DE available to us is stored
in these coefficients. From inequality (43) we find $\vert\,
c_n\vert <\, 0.1\,$. Certainly, first of all, a pending question is
the measurement of $c_0$ and the leading expansion coefficients.
Theoretical predictions, based on connections of $c_n$ with physical
parameters of models ($\varphi_0$, $m$, and others), in no way
discriminate their range of values. Hence, the question hangs solely
on the possibility of experimental assessment of the function
$w(a)$.

We single out three main ways of measuring $w(a)$: structural,
dynamical, and geometrical. Let us consider briefly each of them. We
have already discussed the structural method earlier. Without being
a part of the structure, DE cardinally influences the galaxy
formation rate. It is the structural argument that has led to the
discovery of the DE phenomenon. The achievements of all of
20th-century astronomy established that only a small fraction of
mass enters the large-scale structure of the Universe. The remaining
largest part is, hence, contained in the form of unstructured DE.
Figures 3-5 and 7 depict the growth curves for the seeds of
structure in the case where all $c_n = 0$. For $c_n$ different from
zero, the evolution curves will deform, as can easily be computed
with the help of the GRT. The point, therefore, lies in maximally
precise observational measurements of real growth functions and in
determining the parameters $c_n$ with their help.

How can the growth curves be measured accurately? Unfortunately, one
can hardly rely on the traditional astronomical methods of observing
the compact sources at different wavelengths (stars, including
supernovae, quasars, galaxies, clusters, and so forth) and assessing
based on these data the quantitative characteristics of
inhomogeneous matter distributions in space and time. The obstacle
here is uncontrolled \text{nonlinear} effects of coupling between
density distributions of matter and light. They, eventually, make
impossible the recovery of mass distribution, based on the
luminosity of its baryonic component, with necessary accuracy.

The true breakthrough in the structural method of detection of DE
properties can be expected from dynamical measurements which are
sensitive to gradients of total gravitational potential. An example
can be furnished by statistical measurements of background galaxies
weakly lensed by nearby structures, but even here barriers are met.
The point is that the retrieved surface density of total mass (for a
given redshift), which is the final product of the method, is still
insufficient for a successful comparison with theory. Notably, we
are not in a position to compute the contribution from baryons with
necessary accuracy because of their complex interaction with light
(shock waves, stellar formation, supernovae, dissipation,
ultraviolet background, ionization, cooling, and so on).

In today's discussion of maximally precise methods of measuring DE,
the case in point is largely the exploration of \textit{linear and
quasilinear} structures because, in this case, there is a well
understood and fully controllable theory. An example of a
breakthrough in this field is furnished by measurements of
anisotropy and polarization of cosmic microwave background radiation
(CMBR) that has led to the creation of CSM. A similar breakthrough
is also possible by further developing this model based on accurate
measurements of cosmological parameters, including those of DE.

This breakthrough can be provided by \textit{any} statistical
measurements of large-scale peculiar velocities of matter,
inhomogeneities of gravitational potential, and distributions of
matter density. In the first case, we have to do with the Doppler
effects pertaining to the motion of matter~\cite{bertschinger1990,
kashlinsky2010}, and the measurement of the field of peculiar
velocities based on proper motions of galaxies on the celestial
sphere~\cite{lukash2011}. In the second case, it is the retrieval of
gravitational potential with the help of the weak lensing effect,
but this time at larger scales, where self-crossing of streams of
matter is absent and the contribution from baryons is easily
assessed~\cite{huterer2010}. In the third case, we are dealing with
quantitative distribution of structures as a function of redshifts,
and the detection of baryonic acoustic modulation of the density
perturbation spectrum (see, for example, Ref.~\cite{percival2007}).

Staying on this path, we arrive at the dynamical method of measuring
DE by virtue of the Sachs-Wolfe integral effect. The influence of DE
induces an \textit{additional} source of the CMBR anisotropy,
related to the decay of the linear gravitational field in the
current epoch (see Fig. 5). By measuring this effect, one can
exactly determine the properties of DE, as has already been done
when determining the parameters of the early Universe from
measurements of the anisotropy parameters of CMBR in the
recombination epoch. Quantitatively, here we need to raise the
measurement accuracy only by one order of magnitude.

\begin{figure}[tb]
\epsfxsize=112 mm \centerline{\epsfbox{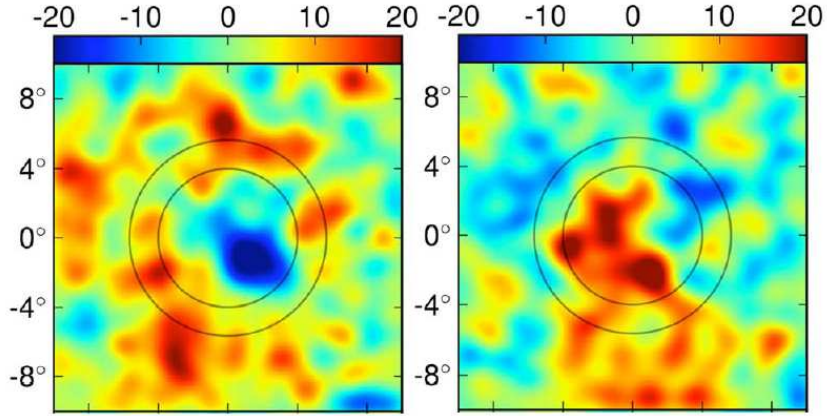}}
\caption{Superimposed portions of a cosmological microwave
background map, which are projected onto 50 voids (a) and 50
superclusters (b) $\sim 100$ Mpc ($4^0$) in size, based on results
of Ref.~\cite{granett2009}. The colorbar of the anisotropy of cosmic
microwave background radiation (the insert at the top) is in
microkelvins.}
\label{50voids}
\end{figure}

The dynamical integral effect predicts the existence of a
cross-correlation between two celestial maps -- those of CMBR
anisotropy, and the large-scale structure of the Universe. Figure 8
shows superimposed regions of CMBR maps, projecting onto 50 voids
(Fig. 8a) and 50 superclusters (Fig. 8b) set apart from the galaxy
distribution in the SDSS catalogue. The mean size of
voids/superclusters is about 100 Mpc. Summation of respective
regions with reduced and increased densities of matter is performed
to amplify the cross-correlation signal against the background of a
random field of primary perturbations. We observe an apparent valley
in the temperature of relic radiation (a cold spot) in directions
towards voids, and a hot spot of the appropriate angular size in the
direction towards superclusters, in full agreement with theoretical
predictions.

These are only the first results which demonstrate the
correspondence with what we already know well from the structural
studies. Here, however, the \textit{precise} cosmological
information is encoded; it can be deciphered in future observations
and used to construct a more precise model of the Universe.
Realization of this wonderful opportunity requires two exact maps of
linear cosmological perturbations: the anisotropy of CMBR, and the
gravitational potential of the large-scale structure. All this is
possible at the already available technological level, but hinges on
building new-generation telescopes, both ground-based and operating
in space.

\section{How to measure the Universe}

Impressive successes in exploring the Universe rely on differential
and dynamical measurements of gradients of gravitational potential,
peculiar velocities, and matter density perturbations by methods of
observational cosmology. There are also, however, direct
astronomical methods of measuring the zeroth-order geometry
bypassing the structure. For brevity, we will call them
\textit{geometrical}. These methods are advantageous at large scales
because the Hubble flows build up with distance, while deviations
from them decrease.

Geometrical tests deal with measurements of distances and times
between events. In the zeroth order, these intervals are controlled
by the function $a(t)\equiv (1+z)^{-1}$ and its derivatives. Both
the properties of DE and other parameters of matter can be
reconstructed by the measured cosmological functions $H(z)$ and
$\gamma(z)$.

There are two classical astronomical methods of determining
geometrical sizes: the measurement of radial distances or angular
scales as functions of $z$. In the first case, we are dealing with
the dependence `apparent magnitude-redshift', $m=m(z)$, and in the
second, the `angular size-redshift', $\theta=\theta(z)$. Other
geometrical tests exist, too (for example, counting the number of
galaxies inside spheres of a given radius; see Fig. 2), but in this
review we consider in more details only the two mentioned above.

The relationship $m=m(z)$ is called the Hubble diagram in astronomy.
It allows determining the scale factor $a(t)$ of the Universe if the
absolute luminosity is known for the objects being analyzed. Indeed,
$\left(a^{-1}\!- 1\right)$ is the redshift of the source, and
$\left(t_0\!-t\right)$ is the distance to it in light years measured
along the light cone of the past. Astronomers frequently use the
\textit{luminosity distance} $r$, which defines the flux $\mathcal
F$ of radiation from a point source of luminosity $L_0$ recorded on
Earth, as if both the source and observer were in the Euclidean
space:
$$
\mathcal F=\frac{L_0}{4\pi r^2}\,.
$$
Instead of $\mathcal F$ and $L_0$, one may use the \textit{apparent}
$m$ and \textit{absolute} $M$ stellar magnitudes, which are related
to the distance $r$ between the source and the observer as
\be
m=M+ 5\left({\rm lg}\, r\left[\mbox{pc}\right]-1\right).
\ee

In order to estimate the absolute magnitude of the source
luminosity, one needs to understand the source structure and the way
it shine, i.e., to have a predictive theory. As a rule, this task is
not solvable with the necessary accuracy, as we are dealing with a
complex gasdynamic nonlinear system with many parameters, be it a
galaxy or cluster of galaxies, a quasar, or a star. For this reason,
one frequently resorts to empirical relationships that offer an
estimate of object luminosity based on its other observable
characteristics. These phenomenological relationships are built and
validated on the nearest sources and then used to calibrate distant
objects of the same class. It is implicitly assumed that near and
distant sources of the same class are alike (the \textit{standard
candle} hypothesis).

While different objects have been proposed to play the role of the
standard candle over years,\footnote{By way of example, the
brightest cD galaxies in centers of rich clusters were used in the
1970-1980s as the standard candle. Currently, supernovae, gamma
bursts, X-ray clusters of galaxies and other bright sources that
yield to observations at large distances pretend on this role.} this
research cannot be used in rigorous cosmology as long as there is no
theory for a system of that complexity, the theory based on
numerical simulations and taking into account numerous, as yet
little-studied, factors which pertain not only to the physical
nature of objects (for example, star formation, nonstationarity,
chemical composition, and nuclear fusion), but also to their
environment, conditions of light propagation, and so on. For such
data, a probability will always exist that we are dealing not with
the effects of geometry or composition of the Universe, but rather
with the internal properties of sources and their evolution
(uncontrolled systematics).

While a standard candle is required for making use of Hubble
diagrams, one needs to know the physical size of the observed object
(the \textit{standard rod} hypothesis) in order to measure distances
by its angular size. Let the physical size of an object be $d$, and
its angular size $\theta$. The \textit{distance based on the angular
size}, $D$, is then defined as
$$
D=\frac{d}{\theta}\,.
$$
Making use of the interval of the Friedmann geometry, we get a
relationship between the distances introduced above: %
\be\label{pere}
r=\left(1+z\right)^2 D=\left(1+z\right)R\,,
\ee
where $R$ is the geodesic comoving distance to the object:
\be\label{Xz}
R\equiv\frac{\bar R}{H_0}=\eta_0-\eta(z)\,,\qquad \bar
R=H_0\int_0\frac{dz}{H}\,.
\ee
($\eta$ is the conformal time). Apparently, the cosmological
function $\bar{R}(z)$ depends only on the properties of matter, not
on $H_0$.

With the help of formula (46), any diagram $m(z)$ can formally be
recast as the dependence $\theta(z)$, and vice versa. This is in no
way surprising, as we are exploring \textit{unified} geometry, even
if resorting to various observational tests. In summary, the recipe
for measuring the Universe is simple: construct diagrams $\theta(z)$
and $\dot\theta(z)$, or $m(z)$ and $\dot m(z)$, or others. If the
physical size of the source or its absolute stellar magnitude are
known, we will manage to determine the distance to it and, hence,
the dimensions of the Universe. This will enable us to reconstruct
the functions $H(z)$, $\gamma(z)$, along with other functions of the
scale factor, and, in turn, determine the composition and properties
of matter and their evolution with time.

\begin{figure}[tb]
\epsfxsize=112 mm \epsfbox{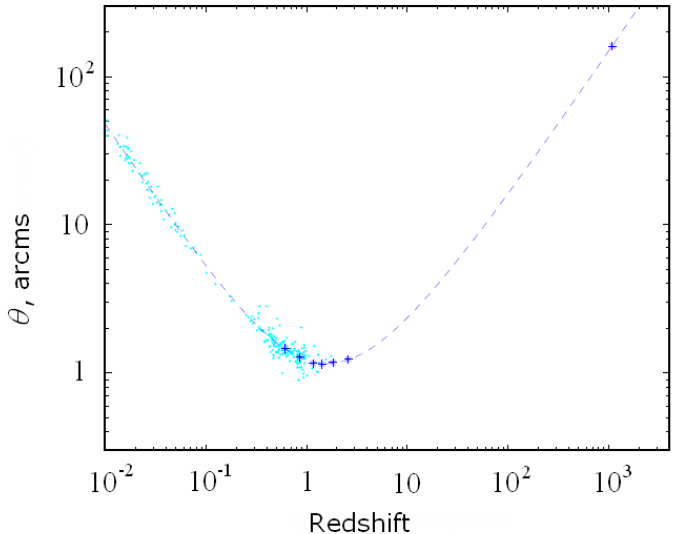}
\caption{Diagram of $\theta(z)$ based on three geometrical tests
reduced to the standard size $d = 9.5$ pc (according to
Ref.~\cite{jackson2006}). } \label{ris_theta}
\end{figure}

The realization of this program requires highly accurate
measurements and knowledge of the nature of the objects under study.
In so doing, these objects can be very diverse. Figure 9 presents a
cumulative diagram of $\theta(z)$ for several geometrical tests
reduced to the standard size of $d=9.5$\, pc. Shown are the results
of measuring distances based on the luminosity of distant Ia
supernovae (points to the left of the minimum), the angular size of
ultracompact radio sources (bases of jets from active galactic
nuclei: six crosses in the lower part of the curve), and the
anisotropy of CMBR (spectrum acoustic oscillations: the cross in the
upper-right part). The dashed line corresponds to the CSM.

The most reliable point of observational data in this diagram is the
right cross determined by the angular size of the acoustic horizon
at the moment of recombination $z_{\rm rec}=1100$. The points and
crosses found from observations of supernovae and ultracompact radio
sources rely, respectively, on the standard candle and size
hypotheses. There are certain physical grounds for using them. In
the first case, we are dealing with the thermonuclear explosion of a
binary white dwarf with total mass in excess of the Chandrasekhar
limit setting the maximum luminosity of a supernova Ia. The second
case pertains to the limiting mass of a central accreting black hole
($\sim 10^{10}\,M_\odot$, $M_\odot$ is the mass of the Sun), which
defines the `standard size' of radio bright quasars. Note, however,
that the available models of supernova explosions and active quasars
do not meet for the moment the requirements placed on accurate
geometrical measurements. Rather, the opposite approach can be
beneficial: by knowing the answer (the parameters of CSM), we can
inquire into the proper cosmological evolution and the nature of
compact sources of that type.

\begin{figure}[tb]
\epsfxsize=112 mm \epsfbox{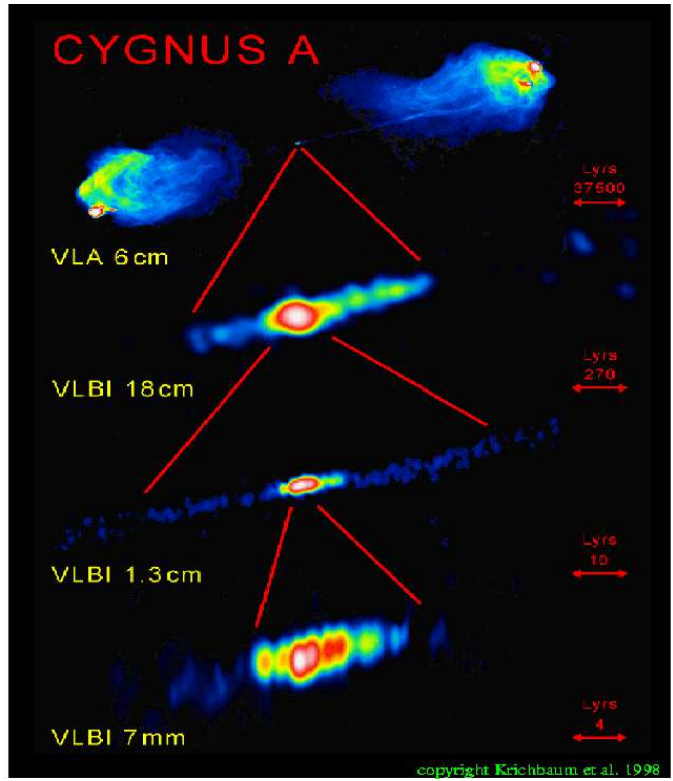}
\vskip5mm \caption{Ultracompact central radio source in Cygnus A
(adapted from Ref.~\cite{krichbaum1998}). $L_{\mbox{yrs}}$ is the
distance expressed in light years.}
\label{ris_ultracom}
\end{figure}

Figure 10 demonstrates the image of an ultracompact radio source
with minimum resolution on the order of a parsec, attained with the
help of a terrestrial radiointerferometric network. The base of a
jet can be seen well, but the internal structure of the source
itself is unresolved. This example illustrates that exploration of
compact objects, which lays the basis for direct geometrical
measurements in the Universe, calls for the development of new
techniques. Their realization is possible by making use of cosmic
radio interferometry.

\begin{figure}[tb]
\epsfxsize=112 mm \epsfbox{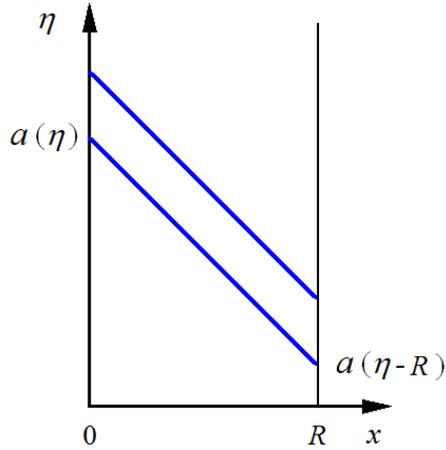}
\caption{Trajectories of light rays in the Friedmann model in
conformal coordinates $(\eta,\,x)$ from the source $(x=R)$ to the
observer $(x=0)$ (see Ref.~\cite{lukash2010}).}
\label{ris_grafik}
\end{figure}

With the help of space telescopes, the interferometric baseline can
be increased and, respectively, its resolution can be improved. Such
instruments not only are capable of measuring distant compact
objects but will also detect time \textit{changes} in their angular
sizes and redshifts, accompanying the expansion of the Universe. The
last two factors are directly linked to the Hubble
function:\footnote{These relationships follow upon differentiating
the functions
$$
\theta=\frac{d}{R\,a(\eta-R)} \qquad \mbox{and} \qquad 1+z =
\frac{a(\eta)}{a(\eta-R)}
$$
over $\eta$ for constant $d$ and $R$, where $\eta$ is the observer
time, and $(\eta - R)$ is the conformal time of the source (Fig.
11).}
\be\label{eef}
\frac{\dot\theta}{\theta}=-\frac{H(z)}{1+z}\,,\qquad \frac{\dot
z}{1+z}=H_0-\frac{H(z)}{1+z}\,,
\ee
This opens new prospects for determining $H_0$ and $H(z)$ at
cosmological distances.

Notice that if the DE density were constant, the redshift of the
sources of the Hubble flow located at $z_0\simeq 2.3$ would not
change: $\dot z(z=z_0)=0$. As this takes place, the redshift of more
distant objects ($z > z_0$) decreases with time, and increases
[$\dot z(z< z_0)>0$] for less distant objects. The boundary $z =
z_0$ is rather sensitive to the evolution of DE. Its detection will
help answering the question of how the DE density varied in the
past.

Determining the evolution effects (48) at cosmological distances
will require several years of monitoring at angular resolution on
the order of microarcseconds, which is technically challenging.
However, even the first cosmic interferometers will be in a position
to measure distances to nearby galaxies by their proper motion
relative to the cosmic microwave background~\cite{lukash2011}.
Moreover, one may measure distances by observing superluminal motion
of the nodes of jets in active galactic nuclei, directed at a small
angle to the line of sight. Such measurements are already being
conducted on Earth (see, e.g., Ref.~\cite{homan2000}) but can be
essentially improved with the help of cosmic interferometers.

\begin{figure}[t]
\epsfxsize=112 mm \epsfbox{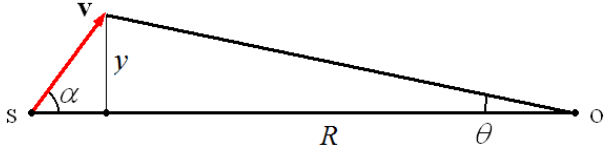}
\caption{The plane $\left( R\,, y\right)$ observer O and the jet
node (the arrowhead) moving with velocity $\mathbf v$ relative to
the source S~\cite{lukash2010}.}
\label{ris_alpha}
\end{figure}

Let us clarify this effect in Fig. 12. Let $\alpha$ be a small angle
between the motion direction of a jet node (the arrowhead) and the
line connecting the observer O with the source S (quasar), and
$\eta$ and $\mathbf{x}=\left( R\,, y\,, 0\right)$ be the conformal
coordinates of the node ($R$ is the geodesic distance from the
observer to the jet; see Eqn (47)). The observation time for the jet
node moving with the intrinsic velocity $\mathbf v$ is given by
$$
d\tau=d\eta-dR=\left(1-{\rm v}\cos\alpha\right)d\eta\,,
$$
and the apparent velocity of its displacement in the observer
tangent plane is defined as
\be
{\rm v}_{\!\bot}=\frac{d\,y}{d\tau}=\frac{{\rm v}\sin\alpha}{1- {\rm
v}\cos\alpha}\, \le \,{\rm v} \Gamma\,.
\ee
This function passes through a maximum at $\cos\alpha=\rm v$, equal
to the jet gamma-factor $\Gamma \equiv\left(1-{\rm
v}^2\right)^{-1/2}$ for the velocity ${\rm v}=|\mathbf v|$
approaching the speed of light.

\begin{figure}[t]
\epsfxsize=112 mm \epsfbox{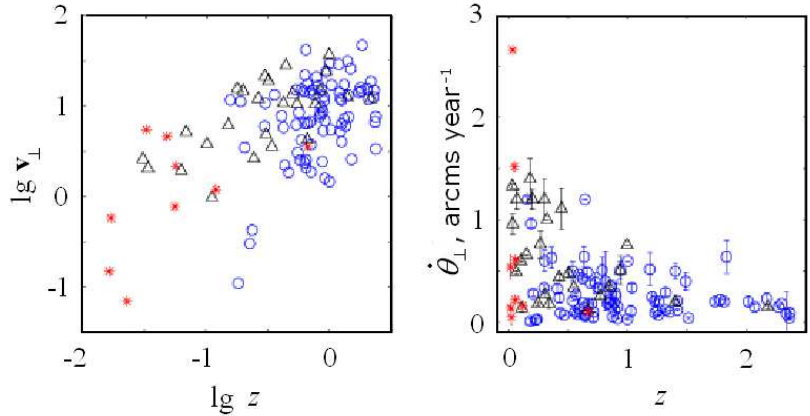}
\caption{Superluminal ejections from quasars (see
Ref.~\cite{zhang2008}): (a) the jet transverse velocity ${\rm
v}_{\!\bot}(z)$, and (b) angular displacement
$\dot\theta_{\!\bot}(z)$.}
\label{ris_vibrosy}
\end{figure}

The jet displacement on the celestial sphere takes then the form
\be\label{ogi}
\dot\theta_{\!\bot} = \frac{{\rm
v}_{\!\bot}}{R}\,\le\,\frac{0.3}{\bar{R}}
\left(\frac{\Gamma}{20}\right)\,\mbox{arcms} \mbox{ year}^{-1}\,,
\ee
which constitutes milliarcseconds per year for the superluminal
velocities of bursts from quasars observed in reality (Fig. 13). By
having extensive statistics of such objects, we will be able to
determine the geometrical characteristics $\bar{R}(z)$ (47) using
the upper envelope of function (50).

We offered several examples illustrating the potential of direct
assessment of geometry and composition of the Universe from
observations of nonlinear objects. They can be augmented by
observations of bright galaxies, compact groups of galaxies, rich
X-ray clusters, gamma bursts, binary quasars, and other active
systems. Exploring the Universe with their assistance remains not as
efficient as that using quasilinear systems and based on
measurements of large-scale structures, microwave background, and
nonrelativistic matter. We may hope that the development of theory
and numerical experiment will make the geometrical methods
competitive in accurately solving the tasks of observational
cosmology.

\newpage

\section{Formation of galaxies in the inhomogeneous Universe}

How do the gravitationally bound halos of matter, reached
hydrostatic equilibrium, emerge out of the quasi-Hubble
flow\footnote{We also call the equilibrium halos relaxed or
virialized systems, with the understanding that it is the
virialization over velocities, not energies.} and how is the mass
distribution of these nonlinear objects linked with the initial
field of density perturbations?

In contrast to the large-scale structure bound up with relatively
small matter density variations and staying at the beginning of its
nonlinear development, the galaxies have evolved from high density
peaks typical for the field of small-scale perturbations (less than
several megaparsecs) and have already passed through the period of
nonlinear relaxation. The reason for that is the amplitude and shape
of the initial density perturbation spectrum which increases toward
short wavelengths (see Ref.~\cite{doroshkevich2012}). This property
of spectrum facilitates early fragmentation of matter into
`halo-blocks' of small mass and their subsequent agglomeration in
more massive halos of galaxies, groups, and clusters.

The hierarchical formation of nonlinear structures from small to
large masses finds support in numerous observations and in numerical
modeling. Important observational argument in favor of the
sequential formation of halo systems is the absence of a large
number of far emission $Ly_\alpha$ lines. The explanation is that
massive systems, in their past, did not undergo a single powerful
burst of star formation involving all halo gas, but formed by way of
subsequent merges of numerous small-mass blocks that passed through
their star formation phases at different times.

By virtue of the existing properties of the spatial S spectrum, the
observable structure of the Universe possesses the following
important features.

--- The formation of galaxies and groups is largely completed, they
contain the dominant part of DM. The process of galactic cluster
formation still continues.

--- Halos and their characteristics are distributed non-uniformly in
space: their number and masses are modulated by the large-scale
structure of the Universe.

--- The first feature offers physical motivation for establishing a
simple linkage between emerging nonlinear halos and peaks in linear
seed density perturbations, based on the model of a quasispherical
collapse [6]. The second feature invites us to apply the
Press-Schechter method to finite-size domains and, in this way, to
couple the mass function of halos with the large-scale structure of
the Universe.

Before discussing theoretical predictions related to the halo mass
functions and their validation against observations and numerical
experiments, we remind the reader of the main characteristics and
elements of the large-scale structure of the Universe. As we have
already stressed, turning to large scales moves us to the initial
stages of structure formation and the first phases of collapse,
which are predominantly one-dimensional. Long-term observations of
the spatial distribution of gravitationally bound virialized DM
halos on scales of up to 300 Mpc demonstrate that the large-scale
structure of the Universe possesses the following typical elements:

--- \textit{filaments} --- linear structures of varying
richness\footnote{The richness of an element of a large-scale
structure is defined by the number of galaxies composing it.} with
lengths of up to tens of megaparsecs, and thickness of several
megaparsecs;

--- \textit{walls} --- flat formations reaching hundreds of megaparsecs over
elongated directions and several megaparsecs in thickness, and
filled with filaments;

--- \textit{nodes} --- rich clusters of galaxies severalmegaparsecs in size
occurring at intersections of filaments;

--- \textit{superclusters} --- extensive domains of space from tens to hundreds
of megaparsecs in size filled with walls, filaments, and nodes;

--- \textit{cosmological voids}\footnote{The term cosmological `void' reflects
the fact that on celestial maps some regions seems to be empty. In
reality, these cosmological empty domains can contain a significant
(yet not dominant) fraction of the Universe's matter, i.e., they are
not really empty. That is why, instead of referring to cosmological
`empty domains', the term `voids' is frequently used.}
--- extensive domains reaching several hundred megaparsecs, where
bright galaxies and galaxy clusters are absent.

The last two structural types are quasilinear because of their large
size. Under the action of gravity, matter moves towards
superclusters which expand more slowly than the mean Hubble flow.
The matter `flows off' the voids in different directions, and their
local volume grows faster than the mean. The mean displacement of
matter elements for the entire history of the Universe is smaller
than the sizes of these structures and makes up 14 Mpc. The voids
occupy the largest part of the Universe (more than 50\%) and their
forms in contours of low density are close to spherical, whereas
superclusters are dense formations with a flattened shape. Both
types of structures can naturally be characterized by the mean
substance density within a given volume. It is smaller than the
Universe's mean value in voids, and larger in superclusters.

Observational assessment of the parameters of voids and
superclusters is ambiguous. The reason for the blurring of voids and
superclusters reconstructed from the spatial light distribution
resides in a nontrivial physical coupling between luminous matter
and the DM composing the backbone of the structure. The resolution
of these formations through the threshold density contrast of the
matter contained in them compared to the mean level is more rigorous
and accurate. We define voids (and superclusters) as
three-dimensional spatial domains with a given negative (positive)
level of matter density contrast. When the absolute value of
threshold contrast decreases, the sizes of structures grow, and
their number decreases. For $\rho < \rho_c$, we obtain a single void
filling the whole Universe.

An important characteristic of voids and superclusters is the mass
function of the halos populating them. Observations indicate that
the cosmological voids nearest to us are impoverished in normal and
dwarf galaxies, but weakly emitting galaxies and gaseous clouds are
encountered. They are observed both through the ultraviolet
absorption lines in the spectra of quasars or bright galaxies placed
behind them and the absorption line of neutral hydrogen (21 cm) in
the spectra of distant radio sources. Galaxies in voids are of
reduced luminosity on average, compared to those in superclusters,
though the slopes of their luminosity functions differ but slightly
from each other~\cite{conroy2005}.

The theory predicts that voids and superclusters contain weak
galaxies of small mass that have not passed the stage of active
stellar formation, as well as primordial objects of the hierarchical
clustering model with masses up to $10^{\,5}\,M_\odot$ and less.
Such objects are incapable of holding gas, so that stellar formation
in them is practically impossible. With growth in measurement
sensitivity, observers are succeeding in finding previously unknown
weak galaxies with masses of $10^7-10^{9}\, M_\odot$ in the Local
Group and the nearest neighborhood~\cite{karachentsev2003}. Their
number agrees well with the predictions of the CSM, yet the question
of finding lessmassive galaxies remains open to date.

The minimum cutoff scale for the spectrum of initial density
perturbations is linked to the physics of DM and is still unknown.
The region of small masses that yields to exploring by means of
observational cosmology (for example, by observing forest of
absorption lines in the spectra of distant quasars) can be linked to
the temperature $T$ of the Universe at an instant of time when the
comoving size $R$ of a given mass $M_R$ coincides with Hubble's
radius of the Universe:
$$
R\simeq 10\left(\frac{10\,\mbox{keV}}{T}\right)\,
\mbox{kpc}\,,\qquad M_R\simeq 10^{\,5}\left(\frac{10\,
\mbox{keV}}{T}\right)^3\,M_\odot\,.
$$

Numerical modeling of processes pertaining to the formation of
structure from cold matter has led to substantial advances in
exploring the mass functions of nonlinear halos in different domains
of the Universe. Starting simulations of the formation process at
$z\sim 20$ with seed masses $M\sim 10^{\,5} \,M_\odot$, to $z\sim 0$
one manages to obtain a spatial distribution of gravitationally
bound systems with typical masses in the range $\sim 10^{12} -
10^{13}\, M_\odot$  (see Refs~\cite{kauffmann1995} -
\cite{bullock2001}), which is similar to the distribution observed.
In this case, all dark matter proves to be involved in the
virialized halos of all sorts of masses, while the mean matter
density inside the halos at the instant of their formation exceeds
the mean density in the Universe by a factor of $\sim 200$. This
corresponds to the predictions of quasispherical collapse.

\section{Mass function of relaxed halos}

The Press-Schechter approximation relies on two assumptions.

(1) The linear perturbation field is Gaussian: the volume fraction
of the Universe where the density contrast $\delta_R(z,\mathbf x)$,
smoothed over a sphere of radius $R$ [see Eqns (61), (62)], exceeds
a certain threshold value $\delta_c$ is given by
\be\label{urr2}
f(M,z) = \frac{1}{\sqrt{2\pi}} \int_\nu^\infty e^{- \nu^{\,\prime
2}\!/2} d\nu^{\,\prime}\,,
\ee
where
\be\label{nun}
\nu=\nu(M,z)\equiv \frac{\delta_c}{\sigma_{\!R}(z)}
\ee
is a monotonically increasing function of argument $M\equiv
M_R=4\pi\rho_{\rm m}R^{\,3}/3$, and
\be\label{frm}
\sigma_{\!R}(z) = \bar{g}(z)\cdot\sigma_{\!R}
\ee
is the variance of density contrast in a sphere of radius $R$ (see
Ref.~\cite{doroshkevich2012} for more detail). The connection of the
normalized factor of density perturbation growth with those
introduced earlier is defined as
\be\label{gzc}
%
\bar{g}\!\left(z\right)=\frac{g\!\left(a\right)}{g\!\left(1\right)}=
\frac{\hat{g}\!\left(a\right)}{\hat{g}\!\left(1\right)}\,.
\ee

(2) The matter in density peaks collapses and stays confined in
gravitationally bound objects: there is such a value $\delta_c$
whereat the mass fraction of matter in the Universe,\footnote{The
numerical coefficient 2 implies that in the course of hierarchical
clustering \textit{all} the cold substance proves to be caught in
virialized halos of all possible masses: $F(>0,z)= 1$ for any $z$.
The argument $M$ is more convenient than $R$ because $M$ is
preserved both for linear perturbations and for halo objects. If the
component of massive neutrinos is included in the matter, we need to
take into account the dependence of growth factor (53) on the
spatial scale.} $F(>M,z)= 2f(M,z)$, turns out to be involved in
virialized halos with individual masses in excess of $M$ by time
moment $z$:
\be\label{urr3}
F(>M,z)=2f(M,z)=\frac{1}{\rho_{\rm m}}
\int\limits_{M}^{\infty}n\,dM\,.
\ee
where
\be
n=n(M,z)\equiv\frac{\rho_{\rm m}}{M}\,{\rm f}_M
\ee
is the differential mass function (the mean number density of halos
with masses $M$ in the interval $dM\sim M$), and
\be
{\rm f}_M={\rm f}_M(z)=\left| \frac{d \ln\sigma_R}{d \ln
M}\right|\,{\rm f}(\nu)
\ee
is the fraction of matter that resides in halos with masses $M$ by
time moment $z$.

A comparison of expressions (51) and (55) allows us to determine the
mean mass function of nonlinear halos in the Universe:
\be\label{fPS}
{\rm f}\!\left(\nu\right) = {\rm f}_{PS} \equiv
\sqrt{\frac{2}{\pi}}\, \nu\, e^{- \nu^{\,2}\!/2}\,.
\ee

The normalization condition of the function  ${\rm f}_M$ is written
down as
\be\label{nku}
\int_0^\infty {\rm f}_M\frac{dM}{M}=1
\ee
and it implies that for any $z$ all the dark matter is confined in
nonlinear halos.We can only inquire about the distribution of these
halos over masses and about the typical halo objects for a given
$z$, containing a dominant part of the Universe's matter. For large
$z$, all the matter is decomposed into small masses, but the share
of large halos grows with time. The mass of typical halos, $M_*$, at
any epoch is defined by the condition
$$
\nu(M_*,z)=1\,.
$$
For the modern Universe, one has $M_* \simeq 10^{13}\,M_{\odot}$.
For small $M$, the function ${\rm f}_M\,\sim \nu\ll 1$. This implies
that halos of small mass are seized by more massive ones in the
course of hierarchical clustering, i.e., they \textit{leave} the set
of $M$ objects and are taken into account only in newly forming
halos of larger mass.

The approximation of linear density contrast in the CSM at the
instant of halo formation in the framework of homogeneous
collapse\footnote{In a simple cycloid model of a collapsing dust
sphere $a= a_0\left(1-\cos(\sqrt{\varkappa}\eta)\right)$, the time
moment $z$ of halo formation coincides with that of collapse,
$t(z)=\int_0 a d\eta= 2\pi a_0/\sqrt\varkappa$. Taking into account
that the radius of a virialized halo is half that of the radius
enclosing the same mass at the moment the cycloid stops, we conclude
that the density of the halo at the instant of its formation exceeds
that of a collapsing sphere at the instant of its stopping by a
factor of 8. Hence, inter alia, a useful estimate follows for the
halo radius at the halo formation instant -- it is approximately six
times smaller than the radius of a sphere enclosing the mass $M$ in
the unperturbed Universe. Speaking figuratively, a halo at its
inception is a fragment of the unperturbed Universe compressed
sixfold in scale.} gives the threshold contrast value
$\delta_c=1.675$, which only weakly depends on variations of
cosmological parameters within their error bounds~\cite{lokas2001}.

Long-term investigations have shown that the results obtained with
the help of this method agree well with the numerical $N$-body
experiment. Nevertheless, the basic formalism does not include
mechanisms of merging and tidal breakup of objects. Hence, as a
rule, the Press-Schechter method is applied to analyzing the spatial
distribution of objects with a sufficiently high mass (such as
clusters of galaxies), for which these effects are insignificant.

In order to apply this method in precise cosmology, account must be
taken of corrections for the nonsphericity of collapse. Analytical
estimates show how the function ${\rm f}(\nu)$ is modified in this
case. For constructing an accurate analytical approximation, we need
to use a more general empirical formula enabling a nonspherical
correction (the Sheth-Tormen approximation~\cite{sheth1999}):
\be\label{fST}
{\rm f}(\nu) = {\rm f}_{ST} \equiv {\rm f}_0\,\sqrt{\frac{2}{\pi}}\,
\tilde\nu\left(1+\tilde\nu^{\,- p}\right) e^{-
\tilde\nu^{\,2}\!/2}\,,
\ee
where $\tilde\nu=\nu/\nu_0$, and find numerically the correction
factors
$$
{\rm f}_0=0.32\,, \qquad \nu_0=1.2 \,,\qquad p=0.6
$$
(${\rm f}_0$ is the normalization coefficient obtained from Eqn
(59)).

\begin{figure}[tb]
\epsfxsize=112 mm \epsfbox{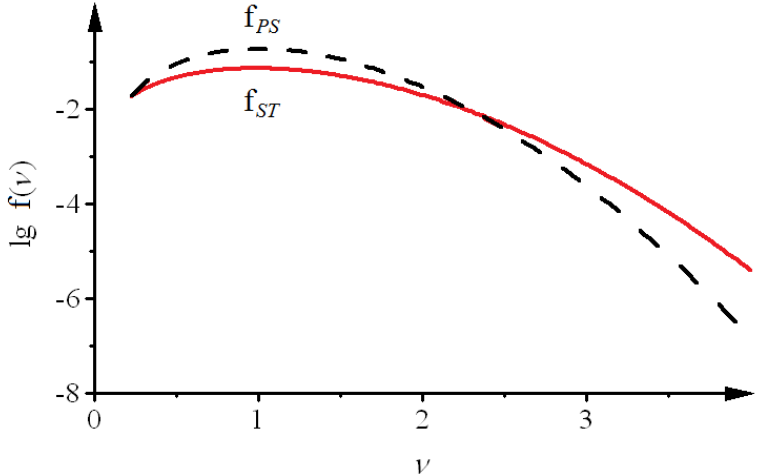}
\caption{Functions ${\rm f}(\nu)$ in the Press-Schechter (PS, dashed
line) and Sheth-Tormen (ST, solid line) approximations of halo mass
functions (see Ref.~\cite{lukash2010}).} \label{graph1}
\end{figure}

Such a corrected approximation, accounting for the Gaussian
perturbation field and the collapse of density peaks, ideally
describes the results of numerical simulations and is widely used in
modern cosmology (Fig. 14). Apparently, the Press-Schechter
approximation somewhat over estimates the number of gravitationally
bound halos with masses near the characteristic $M_*$ ($\nu=1$), but
underestimates the number of massive halos with $M>5\times 10^{14}
\, M_{\odot}$. The intersection of two functions happens at
$\nu=2.3$, which corresponds to the mass of order $\sim
10^{14}\,M_\odot$ in the current epoch (the precise value depends on
the normalization of the spectrum; see Section 14).

In its standard form, the Press-Schechter formalism offers
expressions only for the \textit{mean} mass function of virialized
halos, and misses such large-scale inhomogeneities as voids or
superclusters, in which the local density $\rho_L$ of matter on the
scale $L$ differs from the background one, $\rho_{\rm m}$. In the
next section we will obtain mass functions for halos populating the
inhomogeneous Universe~\cite{arkhipova2007}.

\section{Modulation of galaxies by the large-scale structure}

Local extended domains in the Universe can conveniently be
characterized by the mean matter density inside a sphere of comoving
radius $L$ centered at a point $\mathbf r$:
\be\label{dR1}
\rho_L=\rho_L(\mathbf{r},z)=\int\!\rho\!\left(\mathbf{r}^\prime\!,z
\right)W_L\!\left(|\mathbf{r}-\mathbf{r}^\prime|\right)d\mathbf{r}^\prime
\ee
where
$$
W_L(r)= \frac{3}{4\pi L^3}\left\{
\begin{array}{l}
1\;,\qquad r\le L\\
0\;,\qquad r>L\;\;.
\end{array}
\right.
$$
Let us introduce the parameters of mean density contrast and local
matter density in the domain $L$:
\be\label{skp}
\delta_L\equiv\frac{\rho_L}{\rho_{\rm m}}-1\,,\qquad \Omega_L \equiv
\Omega_{\rm m}\,\frac{\rho_L}{\rho_{\rm m}}\,.
\ee
Adding up expression (62) with $\Omega_{\rm E}$, we obtain the mean
density of matter in the domain $L$. Subtracting from it the
equality $\Omega_{\rm m} + \Omega_{\rm E} = 1$ for a flat Universe,
we obtain the parameter $\Delta_L$ characterizing the mean curvature
of domain $L$:
\be\label{dL}
\Delta_L\equiv\Omega_L -\Omega_{\rm m}\equiv\Omega_{\rm
m}\delta_L\,.
\ee

Distinct from $R$, the free parameter $L$ is assumed to exceed the
inhomogeneity scale ($L > 10$ Mpc), so that spatial domains with
$\rho_L<\rho_{\rm m}$ (voids `v', Fig. 15) and $\rho_L>\rho_{\rm m}$
(superclusters `s') are still in the phase of quasi-Hubble expansion
($|\delta_L|< 1$). As $L \rightarrow \infty$, $\rho_L$ tends to the
mean matter density in the Universe, $\rho_{\rm m}=\rho_{\rm m}(z)$.

In order to construct the local halo mass function we should:

--- resolve the field of linear perturbations of the density
contrast $\delta_R$ into a large-scale background $\delta_L$ and a
small-scale part $\delta_{R\vert L}$ which characterizes the seed
perturbations leading to halo formation in a given spatial domain of
size $L > R$ (see Fig. 15):
\be
\delta_R = \delta_L + \delta_{R|L}\,,
\ee

--- describe a quasispherical collapse on the background $\rho_L$ instead
of the homogeneous background of density $\rho_{\rm m}$.

\begin{figure}[tb]
\epsfxsize=112 mm \epsfbox{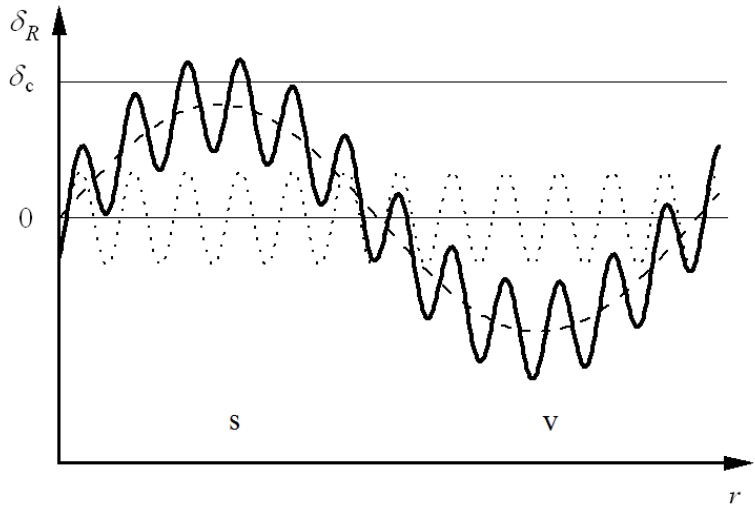}
\caption{Spatial field of density contrast $\delta_R(\mathbf r)$
(solid line) as the sum of long-wave background $\delta_L$ (dashed
line) and short-wave residual $\delta_{R|L}$ (dotted
line)~\cite{arkhipova2007}; $\delta_c$ is the threshold contrast.}
\label{ris83}
\end{figure}

From a physical viewpoint, we are considering the influence of the
local large-scale structure of the Universe on the local process of
hierarchical clustering. Assuming the scale of the structure is
large ($L > R$), its influence on the collapse can be represented in
the form of a power series in spherical harmonics. The monopole term
in this series is the main factor influencing the local statistics
of virialized halos, which lays the basis of this method. The
dynamics of collapse depend only on the mass of the substance
confined within a sphere of radius $R$; however, the regions where
the density contrast exceeds the threshold one are less abundant in
voids than in superclusters. Namely this information is contained in
the background offsetting the Gaussian function $\delta_{R|L}$. At
this stage, we ignore the dipole (density gradient), quadrupole
(tidal forces), and other terms in the structure expansion, which
are linked to the motion of the halo as a whole and the
nonsphericity of collapse. Their influence can be accounted for as
numerical corrections to the halo mass function [see Eqn (60)].

The dispersion $\delta_{R|L}$ of a linear Gaussian field describing
the local properties of density perturbations for $R < L$ takes the
form\footnote{Here, the field crosscorrelation between
$\delta_{R|L}$ and $\delta_L$ is neglected. It is significant only
at $R\sim L$.}
\be\label{urr4}
\sigma^{\,2}_{\!R|L}=\int\limits_{0}^{\infty}P(k)\left\vert
W(kR)-W(kL) \right\vert^2 k^2dk\,.
\ee
Apparently, $\sigma_{\!R|L}\rightarrow 0$ for $M\equiv M_R
\rightarrow M_L$ (the mass in the sphere of radius $L$).

Carrying out computations similar to those in Section 12, we obtain
the halo mass function $M\in (0\,, M_L)$ in the domain
$L$~\cite{arkhipova2007}:
\be\label{1277}
n_L=n_L(M)\equiv\frac{\rho_L}{M}\,{\rm f}_{M|L}\,,
\ee
$$
{\rm f}_{M|L}=\left|\frac{\partial\ln\sigma_{R|L}}{\partial\ln M}
\right|\,{\rm f}(\nu_L)\,,\qquad\int_0^{M_L}{\rm f}_{M|L}\,
\frac{dM}{M}=1\,,
$$
where the function ${\rm f}(\nu)$ coincides with Eqn (58) (or Eqn
(60) in the case of correction for nonsphericity), and $\nu_L$
follows from Eqn (52), with $\sigma_{\!R}$ replaced by
$\sigma_{\!R|L}$, and $\delta_c$ by $\delta_{c|L}$.

The function $\nu_L=\nu_L(M,z)$ grows monotonically with the mass
growth and diverges when $M\rightarrow M_L$. The threshold density
contrast $\delta_{c|L}$ depends on the local density of background.
When computing $\delta_{c|L}$ it should be borne in mind that the
domains of superclusters with $\delta_L>0$ (or voids with
$\delta_L<0$) expand more slowly (faster) than the homogeneous
background. A smaller (larger) threshold contrast $\delta_{c|L}$
with respect to the actual background will assure the collapse of a
given domain of increased density up to a time moment $z$:
\be\label{urr7}
\delta_{c|L}=\delta_c-\delta_L\,.
\ee
This formula is valid in the linear order in $\delta_L$. Numerical
correction is needed to account for nonlinear contributions.

\begin{figure}[tb]
\epsfxsize=112 mm \epsfbox{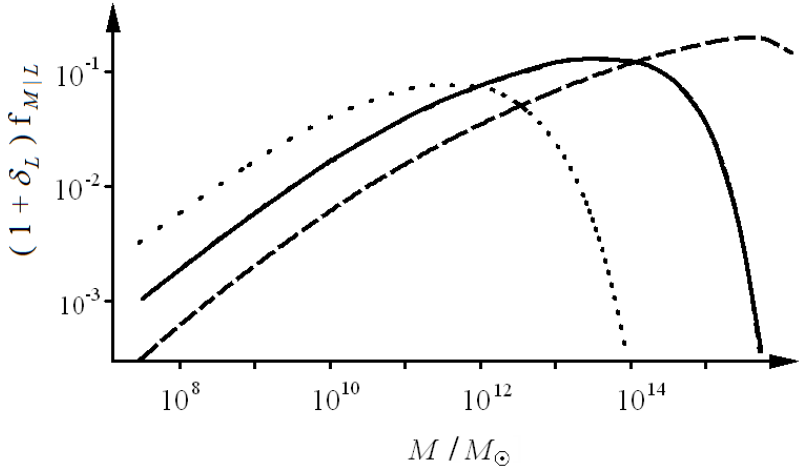}
\caption{The DM fraction $\left(1+\delta_L\right){\rm f}_{M|L}$ in
the halo of masses M (in the interval $\delta M=M$) in the domains
of the modern Universe with the size of $L = 140$ Mpc and density
contrast $\delta_L$ (see Ref.~\cite{arkhipova2007}). The curves
correspond to a supercluster (dashed, $\delta_L=1/2$), flat
background (solid, $\delta_L=0$), and void (dotted, $\delta_L=-
1/2$).} \label{aklm_fig6}
\end{figure}

Figure 16 presents halo mass functions $M n_L/\rho_{\rm m}$ at $z =
0$ in a void, supercluster, and flat domain of space with dimensions
$L = 140$ Mpc and different values of mean density contrast
$\delta_L$. As can be seen, \textit{small-mass} halos form the main
population of zones with a reduced matter density.

For example, the fractions of substance in a void with
$\delta_L=-1/2$, existing in the form of primary isolated blocks --
halos with masses $10^7\,M_\odot$, $10^9\,M_\odot$, and
$10^{11}\,M_\odot$ -- constitute about 1, 10, and 30\%,
respectively. It is easier, therefore, to discover low-mass primary
objects of DM, which are not confined in more massive halo systems,
in voids than in other parts of the Universe.

The solid line for spatially flat domains of the CSM in Fig. 16
characterizes the modern Universe as a whole. We see that the range
$5\cdot 10^{11}-5\cdot 10^{13}\, M_\odot$ of halo masses (massive
galaxies and groups of galaxies) hosts 30\% of the total DM in the
Universe, and massive halos with $M>5\times 10^{14}\,M_\odot$ (rich
clusters of galaxies) include no more than 10\% of its
nonrelativistic matter (${\rm f}_M < 0.1$).

Thus, the process of hierarchical gravitational clustering,
collecting the dominant part of low-mass objects into more massive
systems with $M \lesssim 5\cdot 10^{13}\,M_\odot$  and observed as a
phase of the active merging of galaxies, is completed in the
Universe on the whole. For more massive formations with $M \gtrsim
10^{14} \, M_{\odot}$ this process, seen as the interaction of
groups and clusters, still continues and is far from
completion.\footnote{The patchy structure of most galaxy clusters in
X-ray spectral region offers an additional argument in favor of this
statement.}

If we consider separate domains of space, the coalescence of
galaxies and interaction of galaxies there followed their own ways
in the past and continue to do so. Indeed, clustering of halos in
voids fully ceases with time [see the limit mass $M_L$ in Eqns (65)
and (66)]. In contrast, in `flat' zones and in regions of augmented
density, the clustering and coalescence are progressing and
gradually shift toward larger masses. We may argue that the activity
in the form of coalescences, accompanying the process of galaxy
formation everywhere in the Universe since its beginning,
degenerates with time, continuing in flattened superclusters and
breaking down in voids. The process resembles a burning out fire.

As a result, the morphology of galaxies proves to be different at
various locations in the Universe. Notably, finding \textit{isolated
galaxies} which lack adjacent partners of comparable mass is, for
instance, more probable in voids and close to their boundaries than
in superclusters. In contrast, interacting galaxies are more common
in superclusters, beginning with pairs and ending with large
associations or clusters.

Summarizing, we can conjecture that the theory considering the
formation of the nonlinear structure from DM agrees well with
observations and numerical experiments and has a predictive skill
valuable for further research. There still remain some problems
related to the evolution of the baryon component. Solving them
hinges on further development of computer facilities, observational
techniques, and data processing methods.

The halo mass functions in voids and superclusters differ
substantially. In particular, the spatial density of massive ($M\sim
10^{12}M_\odot$) halos in voids is several times less than in
superclusters, and the distinction in mass functions is even larger
for larger masses. In regions of galaxy clustering, the
observational statistics worsen:
\be
n(M)\simeq 2\cdot 10^{-5}\exp\!
\left(-\frac{M\,}{\,M^*}\right)\,\mbox{Mpc}^{-3},
\ee
where $M^*\simeq 4\cdot 10^{14}\,M_\odot$ on the average in the
Universe. For this reason, the comparison of theory against
experiment relies on \textit{integral} functions of halo masses,
possessing more extensive data statistics.

The integral mass function describes the full number of halos $N_L$
in the domain $L$ (reduced to a unit volume) with masses in excess
of a given one:
\be\label{urr6}
N_L(>M)=\int\limits_{M}^{M_L}n_L(M)\,\frac{dM}{M}\,.
\ee
The differences between the integral functions in voids and
superclusters diverge exponentially with increasing $M$, because the
characteristic values $M^*_L$ of halo masses are different there
(Fig. 17). In the region of small masses (dwarf galaxies), the
integral functions are hardly distinguishable, and the ${\rm
f}_{M|L}$ functions appear more appropriate (see Fig. 16). The
actual distinction between mass populations of voids and
superclusters is stronger because of the nonlinear galaxy
coalescence effects and tidal destruction of galaxies, which partly
suppress the small-mass part of the mass function in superclusters.

\begin{figure}[tb]
\epsfxsize=112 mm \epsfbox{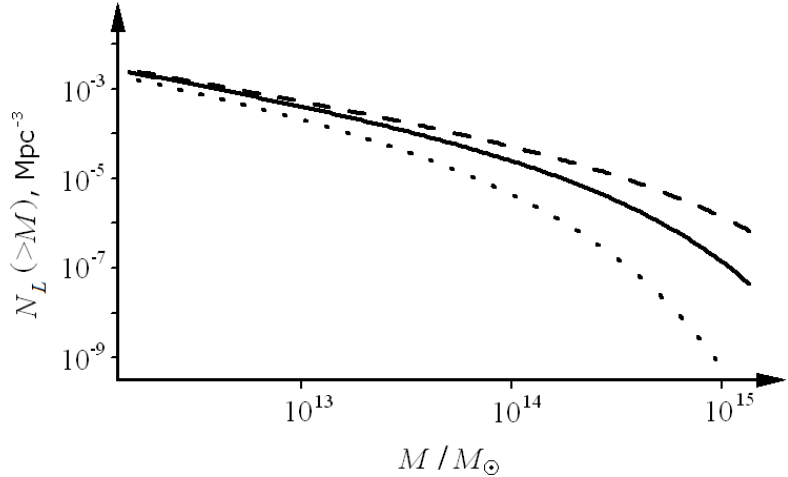}
\caption{ Integral mass functions $N_L(>M)$ of equilibrium halos in
the range $10^{12}\!-10^{15}M_{\odot}$ for the same domains of the
Universe as in Fig. 16.} \label{aklm_fig3}
\end{figure}

The main population of voids is \textit{primordial objects} (POs)
with small masses. The share of the matter in voids in the form of
isolated galaxies with $M < 10^9\,M_\odot$ reaches 10\%, whereas it
does not exceed 1-2\% in other domains of the Universe. Globular
clusters in the central regions of massive galaxies can be stellar
remnants of POs: they could have formed in molecular clouds of the
first small-mass halos and then accumulated in the central regions
of massive galaxies in the process of hierarchical clustering.

The primordial objects in voids can manifest themselves as weak
dwarf galaxies or in the form of $Ly_\alpha$ absorption systems.
Indeed, we can estimate their spatial density and dimension,
assuming that their formation epoch $z_0\sim 10$ and mean mass
$M\sim 10^7\,M_\odot$, as
$$ n_{\rm PO} \sim\frac{\rho_{\rm m}}{M}
\sim 10^4\, \mbox{Mpc}^{-3}, \qquad r_{\rm PO}\sim \frac 16 \,n_{\rm
PO}^{-1/3}\!\left(1+ z_0\right)^{-1}\!\sim \,1 \,\mbox{kpc}\,.
$$
As can be seen, these primary formations are rather loose, so that
lensing effects on them are unimportant. The possibility of
discovering them depends on their subsequent evolution. One may hope
that such objects have retained their gaseous component in voids and
that their column hydrogen density appears sufficient for the
emergence of weak $Ly_\alpha$ absorption lines. The probability of
picking such objects up in the line of sight in a void of size
$L\sim 100$ Mpc is on the order of unity: $n_{\rm PO}r_{\rm
PO}^2L\,\sim 1$.

\section{Normalization of scalar perturbations}

Data on the anisotropy of the CMBR give exact information on the
amplitude and shape of the initial S spectrum in the \textit{early}
Universe and on certain parameters of composition (densities of
nonrelativistic components, and the curvature of space). However, to
trace the transformation of initial inhomogeneities into the
observed galaxies, we also need to know the growth factor for
density perturbations in the \textit{late} Universe, which depends
on the Hubble radius and DE (see Sections 6 and 7). The number and
mass distribution of gravitationally bound relaxed halos, formed
through the development of the gravitational instability of DM,
depend \textit{exponentially} on the amplitude of initial scalar
inhomogeneities in the curvature and growth factor of density
perturbations (see Sections 12 and 13). Owing to this dependence, we
have at our disposal a sensitive test enabling us to determine the
amplitude and shape of the S spectrum together with the most
important supplementary parameters of the Universe's composition
from the quantitative characteristics of the structure.

This test is frequently referred to as the normalization of the
spectrum of cosmological density perturbations. For a given observed
mass distribution of gravitationally bound DM halos, the amplitude
of the power spectrum is a function of cosmological parameters
($\Omega_{\rm m}$, $\Omega_{\rm E}$, and others) within their
variability bounds admitted by the data accuracy.

For historical reasons, the normalization `to sigma 8' -- the
dispersion of density contrast within a sphere of radius $8h^{-1}$
(in this section, $h=H_0/100\,\mbox{km}\,
\mbox{s}^{-1}\mbox{Mpc}^{-1}$), which represents the integral
function of the density perturbation spectrum -- is used most
widely. We denote this dispersion as $\sigma_{11}$ here to emphasize
its actual value: $8h^{-1}\simeq 11$ Mpc. The sphere encloses the
mass of an unperturbed Universe, $M_{11}\simeq 2\cdot
10^{14}\,M_{\odot}$, which is close to that of a typical cluster of
galaxies. This normalization test is therefore ideally oriented to
using observational data on the abundance of galactic clusters.
Moreover, the theoretical analysis of halo mass functions in this
mass range relies on an elaborate analytical formalism (see Sections
12, 13).

The magnitude of dispersion $\sigma_{11}$, which ensures the
observed spatial number density of clusters, essentially depends on
the total matter density $\Omega_{\rm m}$ in the Universe, whereas
changes in all other cosmological parameters affect $\sigma_{11}$
only within 10-20\% of its magnitude. We will demonstrate below the
technique of normalizing the density perturbation spectrum to the
abundance of galaxy clusters at $z = 0$, which satisfies the current
requirements of precise cosmology~\cite{malinovsky2008}.

All cosmological parameters can be spread between two levels
according to their impact on the quantity $\sigma_{11}$:

--- $\Omega_{\rm m}$ (the first level);

--- $\Omega_{\rm E}$, $h$, $n$, ${\rm f}_{\nu}$, $\Omega_{\rm b}$,
$w_{\rm E}$, and others (the second level), where ${\rm
f}_{\nu}\equiv\Omega_\nu/\Omega_{\rm m}$ is the fraction of matter
in the form of massive neutrinos. Among the free parameters of the
second level, the first four are statistically significant (within
intervals of their variability). The rest are fixed for simplicity
($\Omega_{\rm b} h^2 = 0.023$, and $w_{\rm E}=-1$).

A class of models to be utilized represents a rather advanced
variant of the CSM extension at the modern knowledge level: it
includes the nonzero spatial curvature
($\Omega_\varkappa=1-\Omega_{\rm m}- \Omega_{\rm E}$), hot, cold,
and baryonic components of matter ($\Omega_{\rm m} =
\Omega_\nu+\Omega_{\rm M}+\Omega_{\rm b}$), the cosmological
constant, and nonflat (but power-law with the slope $n$) spectra of
the S mode. In this class of models, the exact dependence of the
threshold density contrast $\delta_c$ on $\Omega_{\rm m}$ and
$\Omega_{\rm E}$ in the curved Universe is also taken into
account~\cite{lokas2001}.

The subsequent study is split into two stages. At the first stage,
an optimum value of $\sigma_{11}$ with accompanying individual error
was computed through the comparison of theoretical and observational
differential functions of cluster masses for every realization of
the extended model (with its own set of cosmological
parameters).\footnote{The approach presented here can work with
arbitrary observational data pertaining to virial masses of galaxy
clusters. In the particular case considered here the results of
optical observations of the speeds of galaxies in the nearest 150
clusters with a median value of redshift $z\simeq0.05$ (see
Refs~\cite{girardi1998a, girardi1998b}) were utilized.}

At the second stage, all the computed values of $\sigma_{11}$ (they
are tens of thousands in number, and each has its own error) were
fitted by the approximating dependence of the
form~\cite{malinovsky2008}
\be\label{apr}
\sigma_{11} = \Omega_{\rm m}^{\,A_1+A_2\,\Omega_{\rm m}+A_3\,
\Omega_{\rm E}} \left[ A_4\, +\right.
\ee
$$
\left.+\,A_5\left(\Omega_{\rm m}-A_6\right)
\left(1-A_7\,h-A_8\,n-A_9\,{\rm f}_{\nu} \right) \right],
$$
The optimal values of parameters $A_i \, (i=1,2,\ldots,9)$ were
determined by the Levenberg-Marquardt method of $\chi^2$
minimization. Notice the important distinction of this stage from
the previous one. In the first case, $n(M)$ was a function of a
single variable (mass $M$). In the second case, the function
$\sigma_{11}$ depended on multiple variables ($\Omega_{\rm m}$,
$\Omega_{\rm E}$, $h$, $n$, and $f_{\nu}$).

\begin{figure}[tb]
\epsfxsize=112 mm \epsfbox{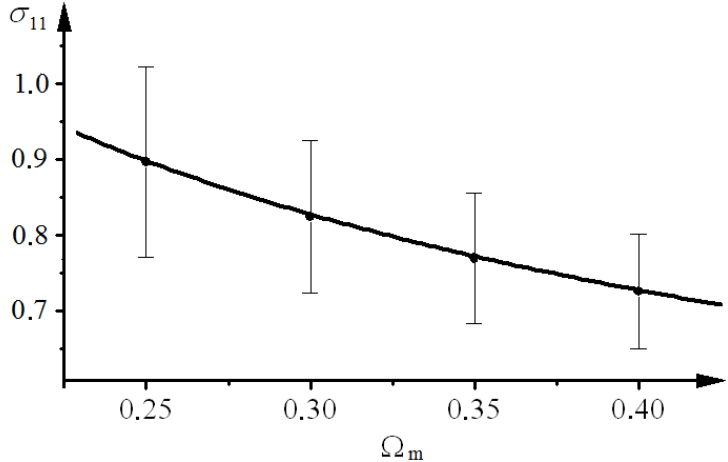}
\caption{The dispersion $\sigma_{11}(\Omega_{\rm m})$ of density
contrast in a sphere of 11 Mpc in radius for the
CSM~\cite{malinovsky2008}. The vertical lines show the accuracy of
observational optical data on the abundance of near clusters of
galaxies (a confidence level of 95\%).} \label{s8-omega}
\end{figure}

Figure 18 illustrates the second phase of normalization. It shows
the slice of the cosmological parameter space of the extended model
in the two-dimensional plane of main parameters ($\sigma_{11}$,
$\Omega_{\rm m}$). In this case, the standard values were taken for
other cosmological parameters ($\Omega_{\rm E} = 1 - \Omega_{\rm
m}=h = 0.7$\,, $n = 1$, $f_{\nu} =0$). The plot confirms the
expected, strong dependence of $\sigma_{11}$ on $\Omega_{\rm m}$.

The experience of working with different data indicates that the
value obtained for $\sigma_{11}$ depends on the selection of
observations. We have already mentioned that this is, unfortunately,
a common story accompanying work with nonlinear objects (clusters of
galaxies in this case). This should be borne in mind when mentioning
the `standard values' of, first of all, such parameters as
$\sigma_{11}$, $H_0$, and $\Omega_{\rm m}$ (or $\Omega_{\rm E}$).
They, to a larger degree than any other quantities, depend on
observations of astronomical objects (stars, quasars, galaxies, and
clusters). There is only one way out: improvement in the quality of
observational data and control of systematic effects related to the
evolution of the baryonic matter component.

\begin{figure}[tb]
\epsfxsize=112 mm \epsfbox{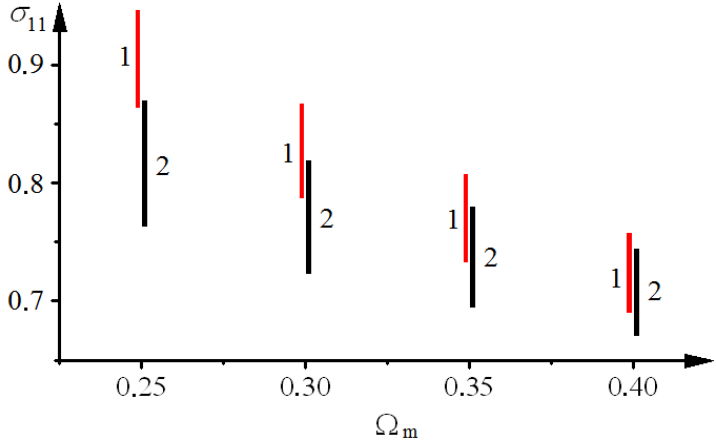}
\caption{Projection of the space of cosmological parameters in the
extended CSM onto the plane ($\sigma_{11}$,$\Omega_{\rm m}$) for the
Sheth-Tormen (bars labeled with 1) and Press-Schechter (labeled with
2) approximations (from Ref.~\cite{malinovsky2008}).}
\end{figure}

We now turn to other parameters. The fact that their contribution to
$\sigma_{11}$ is statistically significant is vividly illustrated,
besides by formulas [see Eqn (71) below], by Fig. 19. All the space
of models is projected there onto the plane ($\sigma_{11}$,
$\Omega_{\rm m}$). As can be seen, the height of bars $\sigma_{11}$
for various values of parameters $\Omega_\varkappa$, $h$, $n$, and
${\rm f}_{\nu}$ is comparable to the size of statistical error, and
exceeds it for some sets of cosmological parameters. This emphasizes
the importance of a thorough account of the contributions from all
model parameters mentioned above.

An interesting result, also seen from Fig. 19, is the increasing
influence of the second-level parameters as the values of
$\Omega_{\rm m}$ reduce. Since the interest of modern cosmology is
focused namely on this domain, this once again confirms the
necessity of simultaneous account of all cosmological parameters
when solving cosmological tasks at a high level of accuracy. The
scatter of $\sigma_{11}$ upon variations of second-level parameters
looks tighter for the Sheth-Tormen approximation, which says
something about its advantage: it ensures the most accurate power
spectrum normalization in work with mass functions of galaxy
clusters (see, for instance, Ref.~\cite{tinker2008}).

We specify the form of formula (70) for the normalization of the
scalar perturbation spectrum as obtained in
Ref.~\cite{malinovsky2008}, which compared the abundance of optical
galaxy clusters with the theoretical halo mass function derived in
the Sheth-Tormen approximation in the extended CSM:
\be\label{apr1}
\sigma_{11}\Omega_{\rm m}^{\,0.54\,+\,0.15\,\Omega_{\rm m}-
\,0.04\,\Omega_{\rm E}}\;-\;\;\;\;
\ee
$$
-0.2\left(\Omega_{\rm m}\!-0.75\right)\left(1-0.2\,h-0.2\,n+
0.8\,f_{\nu}\right)=0.53\pm0.08.
$$

Here, the left-hand side contains combinations of cosmological
parameters, whereas all errors are mapped into the right-hand side
and attributed to the normalization (a confidence level of 95\%).

This example illustrates the specifics of cosmological research with
respect to observational data:

--- the need for \textit{multidimensional} spaces of parameters to determine
the main model characteristics,

--- the need for \textit{numerous} multivariate observational tests to lift
the problem of parameter degeneration.

As follows from the example above, formula (71) alone is
insufficient to normalize the S spectrum amplitude itself: we only
know the relationship between the parameters, and need additional
data to solve the normalization task. Note also an interesting
feature of this formula: it has the form of a `plane' for the
second-level parameters. The coefficients of this plane may be
refined by more complete data of future observations.

\section{Conclusions}

Summing up, we can argue that the standard theory for the structure
formation in the Universe in the framework of the CSM does not
encounter principal problems in explaining the observed matter
motions and distributions. The theory of nonlinear structure
formation hinges on the dominance of DM. It agrees well with
observations and numerical experiments and has predictive skills for
further research. There remain complex problems pertaining to the
nonlinear dynamics of DM and the evolution of the baryonic component
(we consider some of them in Ref.~\cite{doroshkevich2012}). Solving
the problems listed above and those related to them will benefit
from the development of computer facilities, new methods of
observation and data processing, and the refinement of cosmological
parameters aimed at further improvement of the theory and extension
of the CSM.

\subsection*{Acknowledgments.} The authors are indebted to S V
Pilipenko for the fruitful discussions. The work was carried out
with the support of RFBR grants OFI 09-02-12163, 11-02-00857, and
FTsP `Scientific and scientific-pedagogical personnel of innovative
Russia' for 2009-2013 (State contracts P1336 and 16.740.11.0460). A
M M is indebted to the Educational-Scientific complex of FIAN and
the Special Program of the RAS Presidium in support of young
scientists.

\section{Appendices}

\subsection*{A. Quasi-Hubble flows in the GRT} Appendices A, B, and C
present the main features of inhomogeneous cosmological models in
which the Friedmann symmetry is fully broken, yet the departure from
it remains small~\cite{lukash2010}. Such models, deviating only
slightly from the spatially flat Friedmann model, are called here
\textit{weakly inhomogeneous} or \textit{quasi-Friedmann}, the flows
spawning them, quasi-Hubble, and the geometrical departures proper,
considered in the linear order of smallness, cosmological
\textit{perturbations}.

One of the most important geometrical characteristics in the GRT is
the Einstein tensor, or the stress-energy tensor of matter
$T_\mu^\nu$. Energy and momentum are carried in spacetime along
timelike world lines -- flow trajectories. A vector tangent to them,
$u^\mu$, is the eigenvector of tensor $T_\mu^\nu$:
\be\label{tmn}
T_\mu^\nu\, u^\mu = \varepsilon u^\nu\,,
\ee
and its eigenvalue $\varepsilon=\rho c^2$ is the scalar of total
comoving density.\footnote{In the energy units of measurement
adopted by us, energy and mass densities coincide ($c = 1$). Their
different notations $\varepsilon$ and $\rho$, respectively, are only
important for dimensionality recovery and passages to the limit. For
example, $G\rho$ does not contain the physical constant -- speed of
light in vacuum -- and possesses the same dimensionality as $H^2$.}
The unit vector $u^\mu$ ($u_\mu u^\mu = 1$) describes the transport
velocity of total flow energy. By definition, one has
$$
p_{\alpha\nu}\,u^\mu\, T_\mu^\nu = 0\,,
$$
where $p_\mu^{\,\nu}\equiv\delta_\mu^\nu -u_\mu u^\nu$ is the
projective tensor orthogonal to $u^\mu$. Accordingly, the symmetric
tensor $p_{\alpha\nu}T_\beta^\nu$ describes the pressure structure.

In the Friedmann Universe, solution (72) is unique for all forms of
matter: the components of a medium move along the vector $u^\mu$ (in
Friedmann's coordinates $u^\mu=\delta^\mu_0$). The stress-energy
tensor of matter has a universal form which contains the only `new'
scalar $p = p(t)$, the effective pressure in the medium:
\be\label{temi}
T_\mu^\nu=\left(\varepsilon +p\right)u_\mu
u^\nu-p\,\delta_\mu^{\,\nu} =\varepsilon\, u_\mu u^\nu - p
\,p_\mu^{\,\nu}\,.
\ee

For weak perturbations of the Friedmann group, different medium
components move in different directions, but deviations of their
velocities from the mean velocity $u^\mu$ are small. Under this
condition, solution (72) is unique and we obtain the general form of
$T_\mu^\nu$ in the quasi-Friedmann geometry:
\be\label{st6}
T_\mu^\nu= \left(\varepsilon +p\right) u_\mu u^\nu -
p\,\delta_\mu^\nu - s_\mu^{\,\nu}=\varepsilon\,u_\mu u^\nu
-\left(p\, p_\mu^{\,\nu} + s_\mu^{\,\nu} \right),
\ee
which contains, in addition to the form (73), a small stress tensor
$s_\mu^{\,\nu}$ orthogonal to the flow $u^\mu$ ($u^\mu s_\mu^{\,\nu}
=0$). As we see, the total tensor of quasi-Hubble flow pressure
comprises two terms: the diagonal tensor $p$, and the anisotropic
stress tensor $s_\mu^\nu$. The total pressure can be determined by
the trace of the pressure tensor:
\be\label{ptot}
p_{\rm tot}= \frac 13\,p^{\,\mu\nu}T_{\mu\nu}=p + \frac 13\,
s_\nu^\nu\,.
\ee

In a weakly inhomogeneous model, as in the Friedmann model, the
rules of linear superposition are valid for matter components
interacting only gravitationally:
\be\label{pra}
\varepsilon=\Sigma\, \varepsilon_m\,,\qquad p=\Sigma\, p_m\,, \qquad
u^\mu = \Sigma f_m u^\mu_m\,,
\ee
where scalars $f_m$ of partial contributions of matter components
have the form
$$
f_m\equiv\frac{\varepsilon_m+ p_m}{\varepsilon+ p}\,,\qquad \Sigma
f_m = 1\,,\qquad m= 1,2,\ldots,M\,.
$$
For general nonlinear interactions, the resolution of densities and
pressures into components is ambiguous; however, notions of partial
enthalpies $W_m\equiv\varepsilon_m+p_m$ and component velocities
$u^\mu_m$, as well as total $\varepsilon$ and $p$, are preserved.
Notice that these simple linear superpositions of medium component
velocities are valid for terms of zeroth and first orders of
smallness with respect to deviations of velocities $u^\mu_m$ from
the mean one $u^\mu$.

Projecting the Bianchi identities $T_{\mu\,;\,\nu}^\nu=0$ onto
$u^\mu$ and orthogonal directions, we arrive at equations for the
energy and momentum of the total flow:
\be\label{usk0}
u^\nu \varepsilon_{,\,\nu} + 3H_{\rm v}\left(\varepsilon+ p_{\rm
tot}\right)= 0\,,
\ee
\be\label{usk}
a_\mu\equiv u^\nu u_{\mu\,;\,\nu} = u^\nu \left( u_{\mu\,,\,\nu}-
u_{\nu\,,\,\mu}\right)=\frac{ p_\mu^\nu\,p_{,\,\nu}}{\varepsilon +
p}\,,
\ee
where $H_{\rm v}=u^\nu_{\,;\,\nu}/3$ is the Hubble factor of medium
local \textit{volume} expansion, and $a_{\mu}$ is the flow
acceleration ($a_\nu u^\nu= 0$). To make the resolution of pressure
into isotropic and anisotropic parts unique, we demand that the
scalar part of $s_\mu^\nu$ be of zero divergence [see Eqn (81)].

In the class of \textit{quasi-Friedmann} coordinate splits
$(t,\mathbf x)$, where the three-dimensional flow velocity is always
small ($D$ and $v$ are small functions):
\be\label{u}
u_\mu=\left(1+D,\, v_{,\,i}\right),
\ee
the decomposition of geometrical variables into the background and
perturbation parts yields
\be\label{P}
g_{\mu\nu}= g_{\mu\nu}^{(F)}+ h_{\mu\nu}\,,
\ee
\vskip1pt
$$
g_{\mu\nu}^{(F)}={\diag}\left(0\,,\,-a^2\delta_{ij}\right),\qquad
\frac 12\,h_{\mu\nu}=\left(
\begin{array}{cc}
D & \!C_{,\,i}/2  \\
{\rm sym} & \,a^2\!\left(A\,\delta_{ij} + B_{,\,ij}\right)
\end{array}
\right),
$$
\vspace*{3mm}
\be\label{sk}
\varepsilon = \varepsilon^{(F)} + \delta\varepsilon\,,\qquad p =
p^{(F)} + \delta p\,,
\ee
\vskip1pt
$$
s_\mu^\nu ={\diag}\left( 0\,,\,\frac{\Delta S\,\delta_{ij}
-S_{,\,ij}}{8\pi Ga^2}\right),\qquad s_{\mu,\nu}^\nu = 0\,,
$$
where sym denotes a symmetric matrix, and the letter F (discarded
below when possible) labels the Friedmann variables. They are
functions of $t$ and obey the equations
\be\label{82}
\frac 32 H^2 = 4\pi G\varepsilon^{(F)}\,,\qquad
\dot\varepsilon^{(F)}+3H(\varepsilon^{(F)} +p^{(F)})=0\,.
\ee
The scalar sector of perturbations is fully described by four
gravitational ($A$, $B$, $C$, $D$) and four material potentials
($v$, $\delta\varepsilon$, $\delta p$, and $S$ for arbitrary
physical fields!). The scalar $S$ of anisotropic pressure is gauge
invariant, while the remaining seven functions are not.\footnote{It
should be reminded that the vortex and tensor parts of $s_\mu^\nu$
are linked with V and T perturbation modes and are not considered
here. Decompositions (80) and (81) of geometrical objects into a
background and perturbations are not unique: under a small
coordinate transformation $x^\mu \rightarrow x^\mu - \xi^\mu$ we
will obtain a new background (the same background functions, but for
other time $t$) and new perturbations, but the total geometry will
be preserved. Expanding a small arbitrary vector $\xi_\mu=\left(
X,\,a^2 Y_{,i}\right)$ in two potentials $X$ and $Y$, we get the
following gauge transformations for scalar variables:
$$
h_{\mu\nu} \rightarrow h_{\mu\nu} + \xi_{\mu;\nu} + \xi_{\nu;\mu}\,,
\qquad u_\mu \rightarrow u_\mu + X_{,\mu}\,,
$$
$$
A\rightarrow A - H X\,,\qquad B \rightarrow B + Y\,,\qquad C
\rightarrow C + X + a^2\dot Y\,,
$$
$$
D \rightarrow D + \dot X\,,\qquad v \rightarrow v + X\,,\qquad t
\rightarrow t - X\,.
$$
}

The gauge invariant variables for the dimensionless velocity
potential and perturbations of density and pressure in the matter
have the form
\be\label{q}
q = A + H\,v\,,\qquad\delta\equiv \frac{\delta\varepsilon_{\rm c}}{
\varepsilon +p}\,,\qquad \delta_p\equiv\frac{\delta p_{ \rm
c}}{\varepsilon +p}\,.
\ee
where the Lagrangian variables are
\be\label{qec}
\delta\varepsilon_{\rm c}\equiv\delta\varepsilon - \dot\varepsilon\,
v = \varepsilon - \varepsilon_{\rm c}\,,
\ee
\be\label{pv}
\delta p_{\rm c}\equiv\delta p -\dot p\, v = p - p_{\,\rm c} = p_{
\rm tot} - p_{ \rm v}\,.
\ee
The scalar of a volume pressure $p_{ \rm v}=p_{ \rm c}+s_\nu^\nu/3$
and background functions $X_{\rm c}\equiv X^{F}(t_{\rm c})$ describe
distributions on spacelike hypersurfaces of constant
\textit{comoving} time $t_{\rm c}= t+ v$.

The field $q=q(t,\mathbf x)$ plays a central role in the description
of density perturbations~\cite{lukash1980a, lukash1980b}. It has a
double physical sense: on the one hand, it is the dimensionless
potential of the total matter velocity (laboratory interpretation),
while, on the other hand, it is the potential of spatial curvature
(cosmological interpretation). From the Einstein equations in the
lowest order one derives the relationship between the potentials of
the S mode and the field $q$:
\be\label{F}
\frac{\Delta\Phi}{a^2} = 4\pi G\,\delta\varepsilon_{\rm c}\,,\qquad
\Phi=\frac{H}{a}\int \left(\gamma q - S\right)\frac{da}{H}\,,
\ee
\be\label{Eu1}
v - C + a^2\dot B =\frac{\mathfrak q}{H}\,,\qquad \dot v - D=
\frac{\dot q}{H}\,,
\ee
\be\label{Eu}
\delta_p=\frac{\dot q}{H}\,,
\ee
where $\mathfrak{q}\equiv q - \Phi$ is the potential of the peculiar
velocity of matter [see Eqn (101) below]. Hence, notably, having
chosen $v = C = 0$, we obtain the metrics in the Lagrangian
orthogonal reference frame $(t_{\rm c}$,${\mathbf x})$:
\be\label{L}
ds^2=\left(1 - 2\delta_p \right) dt_{\rm c}^2 - \mathfrak{a}^2
\left( \delta_{ij} - 2\mathcal{B}_{,\,ij}\right) dx^i dx^j\,,
\ee
where
$$
u^\mu=\left(1+\delta_p\right)\delta^\mu_0\,,\qquad \mathcal{B}
\equiv \int \mathfrak{q}\,\frac{d \eta}{a^2H}\,,\qquad
\mathfrak{a}\equiv a\cdot\left(1- A\right)= a_{\rm c}\cdot\left( 1-
q\right)
$$
is the scalar scale factor. The proper time $ds$ in geodesic
$\mathbf{x}=const$ is connected to the comoving one, $t_c$, by the
condition $ds=\left(1-\delta_p\right)dt_{\rm c}$.

The key equation (86) is the relativistic Poisson equation and links
the Laplacian of gravity potential $\Phi$ with the comoving density
perturbation $\delta\varepsilon_c$. Equation (88) represents the
relativistic Euler equation, or Newton's second law. It connects the
flow acceleration $\dot q$ with the pressure gradient $\delta p_{c}$
acting on it.

We have four gauge invariant scalars $q$\,,\, $\delta$\,,\,
$\delta_p$\, and $S$, but the four metric potentials $A$, $B$, $C$,
and $D$ are not gauge invariant: any two of them can arbitrarily be
chosen by appropriately selecting the functions $X$ and $Y$ (see
footnote 21). Thus, in total we end with six independent scalars
describing density perturbations in the Friedmann model.
Gravitational equations (86)-(88) impose four constraints on the six
potentials of an S mode. Apparently, the gravity equations alone are
insufficient for describing the dynamics of S perturbations. One
needs information on the physical state of matter in the form of two
missing relations (equations of state). However, even without
imposing constraints on the state of matter, we can derive general
evolution equations for the quasi-Friedmann model, similar to the
Friedmann equations in the homogeneous cosmology.

In order to describe the geometry of weakly inhomogeneous flows, let
us use scalar variables, including \textit{both orders} (the zeroth
and first) of perturbation theory. For the matter, these are the
scalars of total density $\varepsilon$ and pressure $p$, while for
the metric they are the scalar scale factor $\mathfrak a$ and the
Hubble function $H_{\rm v}$ of the volume expansion. Their
relationship is defined as
\be\label{hf}
H_{\rm v} \equiv\frac{\dot b}{b}=\frac 13\, u^\mu_{\,;\mu} = u^\mu
\left(\ln a_{\rm v} \right)_{,\,\mu}
\ee
(the dot above a letter denotes a partial derivative over $t_c$),
where we have introduced the following factors of medium volume
expansion:
\be\label{aeff}
a_{\rm v} \equiv \mathfrak{a}\,{\det}^{1/3}\!\left (\delta_{ij}-
\mathcal{B}_{,\,ij}\right) = b\cdot\left(1 - q\right),\qquad b\equiv
a_{\rm c}\cdot\left( 1- \Delta\mathcal{B}/3\right).
\ee
The generalized Friedmann equation (21) for a weakly inhomogeneous
Universe follows after direct summation of the first equations (82)
and (86) and grouping terms in the function $H_{\rm v}$ and
$\varepsilon$. The other equation (23) follows from Eqn (77) with
account for the relationship $\varepsilon+p_{\rm tot}=(1+\delta_p)
(\varepsilon+p_{\rm v})$.

Consider the structure of quasi-Hubble flow. According to Eqn (89),
the proper distance between two neighboring medium elements,
separated by coordinates $\delta x^i$, is given by
\be\label{com}
\delta r_i=\mathfrak{a} \left(\delta_{ij}-\mathcal{B}_{,\,ij}\right)
\delta x^j\,,\qquad \det\left(\frac{\delta r_i}{\delta x^j}\right) =
a^3_{\rm v}\,.
\ee
Differentiating $\delta r_i$ over the proper time $s$, we obtain the
field of paired velocities of matter motion in the weakly
inhomogeneous Universe:
\be\label{HP}
\delta V_i \equiv \frac{\partial\delta r_i}{\partial s} =
H_{ij}\,\delta r^j\,,\qquad H_{\rm v}=\frac 13 H_i^i\,,
\ee
$$
H_{ij} \equiv H_{\rm c}\,\delta_{ij}-h_{ij}\,,\qquad H_{\rm c}=
\frac{\dot{a}_{\rm c}}{a_{\rm c}}\,,\qquad h_{ij} =\frac{1}{a^2
H}\,\mathfrak{q}_{,\,ij}\,,
$$
where $H_{ij}$ is the matrix of gauge invariant Hubble functions
describing the recession of matter in space. Only one feature here
reminds us of the Hubble expansion: the relative recessional
velocity for points of the medium is proportional to the distance
between them. Yet these velocities are anisotropic and depend on
spatial location.

\subsection*{B. Dynamics of cosmological scalar perturbations} If
motions of quasi-Hubble flow obey Friedmann equation (21), then the
cosmological perturbations behave as oscillators. In order to derive
the evolution equation for the adiabatic scalar $q$, we write the
general relationship between perturbations of comoving pressure and
matter energy density in the following form:
\be\label{oh}
\delta_p=\beta^2\delta +\hat\delta_p\,,
\ee
where the function $\beta^2$ describes the speed squared at which
scalar perturbations propagate in the medium (the speed of sound).
For the Pascal media ($S = 0$), it is as follows:
\be\label{ip0}
\beta^2= \sum_{m, l}f_m\,\beta^2_{ml}\,,
\ee
where $\beta^2_{ml}$ is the acoustic matrix of linear medium
perturbations.\footnote{It connects partial amplitudes of comoving
perturbations of pressure and matter density,
$\delta_p^{(m)}=\beta^2_{ml} \delta^{(l)}_\varepsilon$, and the
following relationships hold true (see Chapter 6 in
book~\cite{lukash2010} for details):
\be
f_m\beta^2_{ml}=f_l\beta^2_{lm}\,,\qquad f_l\,\beta^2 +\bar f_l=
\sum_m f_m\,\beta^2_{ml}\,,\qquad \sum_m \bar f_m = 0\,.
\ee
} The scalar of \textit{isometric} pressure perturbation
$\hat\delta_p$ describes the part of pressure which is not related
to perturbations of the total energy density $\delta$. Indeed, if
initially only adiabatic perturbations are present, then
$\hat\delta_p=0$, and for $\delta=0$, one finds $\delta_p=
\hat\delta_p$. The relationship between $\hat\delta_p$ and field
variables of medium components depends on the equation of state of
matter.

From relationships (88) and (94) it follows that
\be\label{sp}
\delta \varepsilon_c\equiv\left(\varepsilon+p\right)\delta =
\alpha^2 H \left(\dot q - H \hat\delta_p\right),
\ee
$$
\alpha^2 = \frac{\varepsilon+p}{H^2\beta^2} = \frac{\gamma}{4\pi
G\beta^2}\,.
$$
Substituting these formulas into the Poisson equation (86), we
arrive at
$$
\alpha^2a^3\left(\dot q-H\hat\delta_p\right)=\int \alpha^2
\beta^2\Delta q\,\frac{da}{H}\,.
$$
Direct differentiation of this relationship provides the equation
for $q$~\cite{lukash2010}:
\be\label{ur16}
\ddot q + \left(3H+ 2\,\frac{\dot\alpha}{\alpha}\right)\dot q
-\beta^2\frac{\Delta q}{a^2} = I\!\left(\hat\delta_p, S\right),
\ee
$$
I=I\left(\hat\delta_p,
S\right)=\frac{\left(\alpha^2a^3H\,\hat\delta_p
\right)^\cdot}{\alpha^2a^3}-\frac{\Delta S}{4\pi G\alpha^2a^2}\,.
$$
On the left-hand side of equation (98), we have the acoustic
d'Alambert operator for the scalar $q$ describing the general
adiabatic density perturbation. The right-hand side contains the
source of dynamic action of isometric perturbation modes on the
evolution of $q$. For an ideal Pascal media ($S = 0$), a more
compact form of equation follows:
\be\label{ur161}
\frac{\left[\gamma a^2\beta^{-2}\!\left(q^\prime - \bar H\,
\hat\delta_p\right)\right]^\prime}{\gamma a^2}- \Delta q =0\,,
\ee
where the prime stands for the derivative over the conformal time
$\eta$, and $\bar H=aH$.

Equations (98) and (99) are valid for a broad variety of media (in
particular, for fundamental scalar fields). We did not refer to the
information on the microscopic structure of matter. The only
geometric characteristic of a medium needed for the derivation of
key equation (98) is $\beta^2$, the mean velocity squared of
propagation of scalar perturbations in the medium (95). Neglecting
isometric perturbations (for example, in the case of a single
medium, matter and $\Lambda$-term, and others), $\hat\delta_p=0$ and
the equation for $q$ acquires a closed form~\cite{lukash1980a,
lukash1980b}:
\be\label{ur162}
\ddot q + \left(3H+ 2\,\frac{\dot\alpha}{\alpha}\right)\dot q
-\beta^2\frac{\Delta q}{a^2} = 0\,.
\ee

\newpage

\subsection*{C. Eulerian coordinates and the Newton limit} The Eulerian
reference frame $(\tau,\mathbf y)$ is uniquely set by the conditions
$B = C = 0$ and is of interest, in particular, because the
\textit{peculiar} velocity of matter
\be\label{pesk}
\mathbf v_{\rm pec} \equiv - \frac{\mbox{\boldmath $\nabla$}
\upsilon}{a} = -\frac{\mbox{\boldmath $\nabla$} \mathfrak q}{\bar
H}\,,\qquad u_\mu= \left(1+\Psi\,,\,\upsilon_{,\,i}\right),
\ee
where $\upsilon={\mathfrak q}/H$, $\Psi=\Phi - S$, is defined
relative to this grid. From the relationship between Lagrangian and
Eulerian coordinates:
\be\label{tr1}
\tau\equiv t_{E}= t_{\rm c}-\upsilon\,,\qquad {\mathbf y}={\mathbf
x}+{\mathbf S}\,,
\ee
we obtain the metric tensor in the Eulerian representation:
\be\label{eu}
ds^2 = \left(1+ 2\Psi\right) d\tau^2 - \mathfrak{a}^2 d{\mathbf
y}^{\,2}\,,
\ee
where ${\mathbf S}=-\mbox{\boldmath $\nabla$}\mathcal B$ is the
displacement vector of a medium element relative to its initial
position. Apparently, the metric in the Eulerian coordinate system
is independent of the gradients of potential $q$, and the scalar
scale factor
\be\label{eu1}
\mathfrak{a}= a(\tau)\cdot\left(1-\Phi\right)
\ee
fully describes the locally isotropic observer space filled with
inhomogeneousmatter. Metric (103) is the relativistic limit of the
weak field in the Friedmann model, and for the nonrelativistic
substance ($\vert S/\Phi\vert\ll 1$) we obtain the Newtonian limit
$\Psi=\Phi$.

Since the curvature scalar and gravitational potential are small
($q\sim\Phi\lesssim 10^{-5}$) in the real Universe within the
observed structure scale, we can drop terms $\upsilon, \Psi$, and
$\Phi$ in Eqns (102)-(104), respectively, and introduce the physical
Eulerian coordinate for the position of medium points ${\mathbf
r}\simeq a\,\mathbf y$. Hence we obtain a convenient approximation
for the description of the quasilinear stage of structure formation
in the Universe:
\be\label{az}
{\mathbf r}=\left(1+z\right)^{-1}\!\left({\mathbf x}+{\mathbf
S}\right),\qquad \mathbf{v}_{\rm pec}=\dot{\mathbf r}-H{\mathbf r}=
a\,\dot{\mathbf S}\,.
\ee
Under the additional assumption of the smallness of the speed of
sound, from equation (100) it follows that $q\simeq
q_0=q_0(\mathbf{x})$. Substituting it into formula (105), we obtain
the Zel'dovich approximation (see Ref.~\cite{doroshkevich2012} for
more detail):
\be\label{az0}
{\mathbf S}=-g\,\mbox{\boldmath $\nabla$} q_0\,,\quad
\mathbf{v}_{\rm pec} =-\nu\,\mbox{\boldmath $\nabla$} q_0\,,
\quad\delta= g\,\Delta q_0\,,\quad \Phi=\phi\,q_0\,,
\ee
where the growth factors are introduced [see Eqn (12)]. They depend
only on time and the relationship $a\phi/g=3\Omega_{\rm
m}H_0^2/2=const$ holds true.

\end{document}